\documentclass[usenatbib]{mnras}
\usepackage{lmodern}

\usepackage[T1]{fontenc}
\usepackage[utf8]{inputenc}
\usepackage{color}
\usepackage{amsmath}
\usepackage{amssymb}
\usepackage{graphicx}
\graphicspath{{figures/}}
\usepackage{esint}
\usepackage{multirow}
\usepackage[normalem]{ulem}

\AtBeginDocument{\hypersetup{
linkcolor=linkblue,
citecolor=linkorange,
urlcolor=urlblue}}

\providecommand{\eprint}[1]{\href{http://arxiv.org/abs/#1}{#1}}
\providecommand{\adsurl}[1]{\href{#1}{ADS}}

\usepackage{ifthen}\def\eprinttmp@#1arXiv:#2 [#3]#4@{\ifthenelse{\equal{#3}{x}}{\href{http://arxiv.org/abs/#1}{#1}}{\href{http://arxiv.org/abs/#2}{arXiv:#2} [#3]}}
\renewcommand{\eprint}[1]{\eprinttmp@#1arXiv: [x]@}

\definecolor{urlblue}{rgb}{0,0,0.9}
\definecolor{linkblue}{rgb}{0,0,.8}
\definecolor{linkgreen}{rgb}{0,0.45,0}
\definecolor{linkpurple}{rgb}{0.7,0.0,0.4}
\definecolor{linkorange}{rgb}{0.7,0.1,0.0}

\makeatletter

\pdfpageheight\paperheight
\pdfpagewidth\paperwidth

\definecolor{somegreen}{cmyk}{0,0.49,0.98,0.09}
\definecolor{red}{rgb}{1,0,0}
\definecolor{magenta}{cmyk}{0,1,0,0}
\definecolor{lavender}{cmyk}{0.16,0.67,0,0.57}
\definecolor{darkgreen}{rgb}{0,0.65,0.05}
\definecolor{antiquefuchsia}{rgb}{0.33, 0.1, 0.89}

\let\jnl@style=\rm
\def\ref@jnl#1{{\jnl@style#1}}

\def\aj{\ref@jnl{AJ}}                   
\def\actaa{\ref@jnl{Acta Astron.}}      
\def\araa{\ref@jnl{ARA\&A}}             
\def\apj{\ref@jnl{ApJ}}                 
\def\apjl{\ref@jnl{ApJ}}                
\def\apjs{\ref@jnl{ApJS}}               
\def\ao{\ref@jnl{Appl.~Opt.}}           
\def\apss{\ref@jnl{Ap\&SS}}             
\def\aap{\ref@jnl{A\&A}}                
\def\aapr{\ref@jnl{A\&A~Rev.}}          
\def\aaps{\ref@jnl{A\&AS}}              
\def\azh{\ref@jnl{AZh}}                 
\def\baas{\ref@jnl{BAAS}}               
\def\bac{\ref@jnl{Bull. astr. Inst. Czechosl.}}
\def\caa{\ref@jnl{Chinese Astron. Astrophys.}}
\def\cjaa{\ref@jnl{Chinese J. Astron. Astrophys.}}
\def\icarus{\ref@jnl{Icarus}}           
\def\jcap{\ref@jnl{J. Cosmology Astropart. Phys.}}
\def\jrasc{\ref@jnl{JRASC}}             
\def\memras{\ref@jnl{MmRAS}}            
\def\mnras{\ref@jnl{MNRAS}}             
\def\na{\ref@jnl{New A}}                
\def\nar{\ref@jnl{New A Rev.}}          
\def\pra{\ref@jnl{Phys.~Rev.~A}}        
\def\prb{\ref@jnl{Phys.~Rev.~B}}        
\def\prc{\ref@jnl{Phys.~Rev.~C}}        
\def\prd{\ref@jnl{Phys.~Rev.~D}}        
\def\pre{\ref@jnl{Phys.~Rev.~E}}        
\def\prl{\ref@jnl{Phys.~Rev.~Lett.}}    
\def\pasa{\ref@jnl{PASA}}               
\def\pasp{\ref@jnl{PASP}}               
\def\pasj{\ref@jnl{PASJ}}               
\def\rmxaa{\ref@jnl{Rev. Mexicana Astron. Astrofis.}}%
\def\qjras{\ref@jnl{QJRAS}}             
\def\skytel{\ref@jnl{S\&T}}             
\def\solphys{\ref@jnl{Sol.~Phys.}}      
\def\sovast{\ref@jnl{Soviet~Ast.}}      
\def\ssr{\ref@jnl{Space~Sci.~Rev.}}     
\def\zap{\ref@jnl{ZAp}}                 
\def\nat{\ref@jnl{Nature}}              
\def\iaucirc{\ref@jnl{IAU~Circ.}}       
\def\aplett{\ref@jnl{Astrophys.~Lett.}} 
\def\apspr{\ref@jnl{Astrophys.~Space~Phys.~Res.}}
\def\bain{\ref@jnl{Bull.~Astron.~Inst.~Netherlands}}
\def\fcp{\ref@jnl{Fund.~Cosmic~Phys.}}  
\def\gca{\ref@jnl{Geochim.~Cosmochim.~Acta}}   
\def\grl{\ref@jnl{Geophys.~Res.~Lett.}} 
\def\jcp{\ref@jnl{J.~Chem.~Phys.}}      
\def\jgr{\ref@jnl{J.~Geophys.~Res.}}    
\def\jqsrt{\ref@jnl{J.~Quant.~Spec.~Radiat.~Transf.}}
\def\memsai{\ref@jnl{Mem.~Soc.~Astron.~Italiana}}
\def\nphysa{\ref@jnl{Nucl.~Phys.~A}}   
\def\physrep{\ref@jnl{Phys.~Rep.}}   
\def\physscr{\ref@jnl{Phys.~Scr}}   
\def\planss{\ref@jnl{Planet.~Space~Sci.}}   
\def\procspie{\ref@jnl{Proc.~SPIE}}   

\newcommand{\dd}{\textrm{d}}

\definecolor{darkred}{rgb}{0.15, 0.15, 0.9}

\newcommand{\one}{$1\times 2$pt}
\newcommand{\three}{$3\times 2$pt}
\newcommand{\six}{\ensuremath{6\times 2}pt}

\definecolor{darkgreen}{rgb}{0.0, 0.4, 0.0}

\makeatother

\title[The $6\times2$pt method]{The 6x2pt method: supernova velocities meet multiple tracers}

\author[Quartin, Amendola \& Moraes]{Miguel Quartin$^{1,2,3}$, Luca Amendola$^{3}$ and Bruno Moraes$^{1}$ \\
$^{1}$Instituto de Física, Universidade Federal do Rio de Janeiro,
21941-972, Rio de Janeiro, RJ, Brazil\\
$^{2}$Observatório do Valongo, Universidade Federal do Rio de Janeiro,
20080-090, Rio de Janeiro, RJ, Brazil\\
$^{3}$Institute of Theoretical Physics, Heidelberg University,
Philosophenweg 16, 69120 Heidelberg, Germany
}

\date{\today }

\begin{document}
\label{firstpage}
\pagerange{\pageref{firstpage}--\pageref{lastpage}}

\maketitle

\begin{abstract}
   We present a new methodology to analyse in a comprehensive way large-scale and supernovae (or any other distance indicator) surveys. Our approach combines galaxy and supernova position and redshift data with supernova peculiar velocities, obtained through their magnitude scatter, to construct a \six\ analysis which includes six power spectra.    The 3$\times$3 correlation matrix of these spectra expresses exhaustively the information content of the surveys at the linear level. We proceed then to forecast the performance of future surveys like LSST and 4MOST with a Fisher Matrix analysis, adopting both a model-dependent and a model-independent  approach. We compare the performance of the \six\ approach to the traditional one using only galaxy clustering and some recently proposed combinations of galaxy and supernovae data and quantify the possible gains by optimally extracting the linear information. We show that the \six\ method shrinks the uncertainty area in the $\sigma_8, \, \gamma$ plane by more than half when compared to the traditional method.    The combined clustering and velocity data  on the growth of structures has uncertainties at similar levels to those of the CMB but exhibit  orthogonal degeneracies, and the combined constraints yield improvements of factors of 5 or more in \emph{each} of the five cosmological parameters here considered.  Concerning the model-independent results, we find that our method can improve the constraints  on $H(z)/H_0$ in all redshift bins by more than 70\%  with respect to the galaxy clustering alone and by 30\% when supernova velocities (but not clustering) are considered, reaching a precision of 3--4\% at high redshifts.
\end{abstract}


\maketitle

\section{Introduction}

Large-scale cosmological surveys are providing ever more important constraints on the evolution of our Universe and its composition. Recent results from surveys like eBOSS~\citep{eBOSS:2020yzd}, KiDS~\citep{KiDS:2020suj} and DES~\citep{DES:2021wwk} already tightly constrain the standard cosmological model, $\Lambda$CDM, and some of its variants. Future surveys like DESI~\citep{desicollaboration2016}, J-PAS~\citep{Bonoli:2020ciz}, Euclid~\citep{Laureijs:2011:1,Amendola:2016saw} and the Rubin Observatory Legacy Survey of Space and Time (LSST)~\citep{LSSTScience:2009jmu}
will provide deeper datasets and the uncertainties on most cosmological parameters are expected to reach and even surpass the 1\% precision level.  On the other hand, there is currently in the cosmological community a large effort to test the accuracy of these observations, especially because of the present tension between measurements of $H_0$ from the CMB~\citep{Planck:2018vyg} and from the astronomical distance ladder using parallaxes, Cepheids and type Ia supernovae (SN)~\citep{Riess:2020fzl}.

The distribution and growth of structures in the universe is usually described using the density perturbation amplitude in scales of 8 Mpc$/h$ (through the parameter $\sigma_8$) and the linear growth rate
\begin{equation}
    f \equiv \frac{\dd\log\delta}{\dd\log a} = -\frac{\dd\log D_+(z)}{\dd\log (1+z)} \simeq \Omega_{m}(z)^\gamma \,.
\end{equation}
Here $\delta$ is the dark matter density contrast, $D_+(z)$ is the linear growth factor and  the growth rate index $\gamma$, assumed constant, represents a simple parametrisation often used to account for the growth of structures in modified gravity models, its value in General Relativity being $\gamma = 0.545$ \citep{Peebles1980,Amendola_2004, Linder2005, LinderCahn2007}.
Another frequently used alternative is to measure the quantity $f(z) \sigma_8(z)$ in different redshift bins \citep{SongPercival2009, PercivalWhite2009, WhiteSongPercival2009}.

Most of these surveys were proposed to focus on redshift and image catalogs of galaxies in order to analyse their clustering and weak lensing properties. However, more recently it was realised that these measurements of the large scale structure (LSS) of the universe can be combined with data from type Ia supernova explosions to great benefit~\citep{Castro:2015rrx,Howlett:2017asw,Garcia:2019ita,Amendola:2019lvy,Graziani:2020kkr}. The scatter of SN around the mean depends on their peculiar velocity field, which, in turn, is correlated to the density contrast in linear perturbation theory~\citep{hui2006,Gordon:2007zw}. This makes SN a very useful tracer of the velocity fields of the universe, and both velocity and density tracers are very complementary mainly because they have different degeneracies with the linear bias. In particular, the 2-point velocity-velocity correlation function is independent of the linear biases which relate dark matter to each class of baryonic tracer.

The method can be straightforwardly applied to any form of cosmological standards such as standard rulers, standard clocks, and standard sirens.

The analysis of supernova clustering extends their utility beyond their more traditional use in a Hubble diagram simply as measurements of distances. In fact, the idea of extracting further information of SN in general has received more attention in the last decade, mostly due to the prospect of increasing the number of detected SN by around two orders of magnitude with the upcoming LSST survey. Besides their use as peculiar velocity tracers mentioned above, they have also been considered and used as probes of gravitational lensing, both in the weak~\citep{Quartin:2013moa,Castro:2014oja,Scovacricchi:2016ylt,Macaulay:2016uwy,DES:2020kbf} and strong regimes~\citep{Grillo:2018ume,Grillo:2020yvj}.

\citet{Johnson:2014kaa} and \citet{Castro:2015rrx} discussed in detail methodologies to extract peculiar velocity information from supernova data and obtained constraints on clustering without the need to employ any galaxy data.  \citet{Johnson:2014kaa} made use of nearby supernova data and measured $\sigma_8 = 0.86 \pm 0.18$. \citet{Castro:2015rrx} analysed the JLA supernova catalog~\citep{Betoule:2014frx} and, combining the SN velocity and SN lensing observables, obtained a joint measurement of $\sigma_8 = 0.65^{+0.23}_{-0.37}$ and the growth rate index $\gamma = 1.38^{+1.7}_{-0.65}$. SN velocities were also used to measure $f \sigma_8$ at low redshifts, where the dependence on cosmology is weak, by~\cite{Huterer:2016uyq,Boruah:2019icj}. The joint use of galaxy and SN clustering was first proposed in more detail by~\citet{Howlett:2017asw}, where Fisher Matrix (henceforth FM) forecasts were performed for measuring $f\sigma_8$. Similar forecasts were also performed by~\cite{Palmese:2020kxn} combining future galaxy and standard siren data. More recently, the multi-tracer prospects of combining SN and siren data was also investigated by~\cite{Libanore:2021jqv} and a first measurement of the clustering of both core colapse and type Ia SN from the Zwicky Transient Facility was performed by \citet{Tsaprazi2021}.

\citet{Amendola:2019lvy} recently showed that one can employ the clustering of standard candles to measure the normalized expansion rate $H(z)/H_{0}$ without assuming neither a specific background, nor a linearly perturbed cosmological model, nor a bias model. The method was dubbed Clustering of Standard Candles (CSC) and it provides therefore interesting constraints that are, to a large extent, model independent. The CSC method as originally proposed makes no use of galaxy data and employs only standard candles, like SN, exploiting both their luminosity distance and their clustering properties.

In this work, we show that both the CSC method above and the one proposed by~\citet{Howlett:2017asw} can be extended and improved by taking into account both galaxies and supernova as density tracers while retaining the latter also as peculiar velocity tracers. The method assumes one has a survey of galaxies and of standard candles that overlap in some redshift range. We will also assume for simplicity that the surveys  overlap in some sky area, but one can extend the method even to non-overlapping sky regions. The standard candles might be in part hosted by the galaxies of the galaxy survey, but the results are only very weakly dependent on this. From the galaxy redshift survey one can extract the density contrast $\delta_{\rm g}(k,z)$ in Fourier space. From the standard candle survey, one extracts the density contrast $\delta_{\rm s}(k,z)$ and, by exploiting the scatter of the apparent magnitude, the radial peculiar velocity field $v_{\rm s}(k,z)$. As long as we stay in the linear regime, we can assume that these three random fields $\{\delta_{\rm g}$, $\delta_{\rm s}$, $v_{\rm s}\}$, obey a Gaussian multivariate statistic with covariance matrix formed by six two-point auto and cross-correlation functions: hence the name CSC--$6\times2$pt we assign to this method, in analogy to the $3\times2$pt method employed to combine shear lensing and galaxy clustering. The CSC--$6\times2$pt method combines the advantages of multiple tracers and of the clustering of standard candles.\footnote{Forecasts for multi-tracer techniques combining density and velocity tracers in the context of peculiar velocity surveys using not supernovae but only galaxy data (Fundamental Plane and Tully Fisher relations) were also performed by~\cite{Howlett:2016urc}.
}
We will work in Fourier space throughout this manuscript, so the correlation functions are represented by the corresponding power spectra.

We derive the Fisher Matrix from this multi-dimensional Gaussian and forecast constraints from future surveys. We take two different approaches. The first is a model-dependent one and the most often used in cosmology, in which we make use of traditional $\Lambda$CDM parameters in addition to the growth-rate index $\gamma$ and of a number of nuisance parameters to account for the galaxy and supernova bias and the non-linear velocity smoothing factors. Our main focus will be on forecasts of joint constraints of the parameters $\{\gamma,\,\sigma_8\}$ for two reasons: first, these are perturbative quantities for which there is still much more model-dependence than background parameters such as $\Omega_{m0}$ and, second, there is currently a moderate tension on $\sigma_8$ measurements from CMB~\citep{Planck:2018vyg} and measurements of galaxy clustering and weak-lensing~\citep{KiDS:2020suj,DES:2021wwk}. This tension is robust even in non-flat models or models with large neutrino masses, but can be alleviated in $w$CDM models~\citep{KiDS:2020ghu}. We will nevertheless also discuss the constraints in $\Omega_{m0}$, $\Omega_{k0}$ and $h$ that can be achieved with the \six\ method alone as well as in combination with the CMB.

The second approach makes use of the model-independent method of~\cite{Amendola:2019lvy} to put constraints on the Hubble function $H(z)/H_0$ in each redshift bin, as well as on the matter power spectrum and bias functions at different redshift and $k$ bins. The constraints are in this case independent of the bias and of the cosmological model at the background and linearly perturbed level, so we do not need any parametrization. Although the method can be directly extended to  non-flat spaces, here we will assume that the spatial curvature is negligible. Direct constraints on $H_0$ are not possible with SN alone due to its complete degeneracy with their intrinsic magnitude, but could be achieved with standard sirens. Because of the $z$ and $k$ binning, the number of free quantities to constrain is much larger in the model-independent case, but competitive bounds can still be obtained. In general, model-dependent approaches typically results in stronger constraints, but specific to the chosen model or parametrization.

\section{The covariance and fisher matrices}

A standard candle with radial peculiar velocity $v_{\rm s}$ (in units of the speed of light) induces a change in the luminosity distance $D_{L}$ given by~\citep{hui2006,Davis:2010jq}
\begin{equation}
    \frac{\delta{D_{L}}}{D_{L}}=v_{\rm s}\left[2-\frac{d\log D_{L}}{d\log(1+z)}\right]\,.
\end{equation}
In turn, a small change in $D_{L}$ induces a change $\delta m$ in the apparent magnitude,
\begin{equation}
    \frac{\delta D_{L}}{D_{L}}=\frac{\log10}{5}\delta m \,.
\end{equation}
Therefore, the radial peculiar velocity of a standard candle is related to the scatter $\delta m$ of its apparent magnitude as
\begin{equation}
    v_{\rm s}=\frac{\log10}{5}\delta m\left[2-\frac{d\log D_{L}}{d\log(1+z)}\right]^{-1}\label{eq:magn-vel} \,.
\end{equation}
In this way, the peculiar velocity field of standard candles can be obtained via their magnitude scatter.

The statistical uncertainty $\sigma_{{\rm int}}$ in the magnitude of a SN is associated via the distance modulus relation to an uncertainty in the redshift~\citep{Amendola:2019lvy}:
\begin{equation}
    \sigma_{v,{\rm eff}}^{2}\!\equiv\!\left[\frac{\log10}{5}\sigma_{{\rm int}}\right]^{2}\!\left[2-\frac{d\log D_{L}}{d\log(1+z)}\right]^{-2}\!\!\!+\frac{\sigma_{v{\rm ,nonlin}}^{2}}{c^{2}}.\label{eq:new-sv-1}
\end{equation}
There is an extra scatter of around $0.05z$ mag due to lensing~\citep{Jonsson:2010wx,Quartin:2013moa} but this is negligible for our purposes. Systematic errors in distances have also been assumed to be small by~\cite{Howlett:2017asw}.

Let us now consider three Gaussian fields in Fourier space with zero mean: the density contrast $\delta_{\rm s}$ of the standard candles (from now on we refer specifically to supernovae Ia), their peculiar velocity field $v_{\rm s}$, and the galaxy density contrast $\delta_{\rm g}$. A fraction $\phi$ of the supernovae could be hosted by one of the galaxies in the sample. Although the expected fraction depends considerably on the survey strategy, in practice this quantity has negligible impact on our forecasts.  We also introduce the linear bias for each species, $b_{\rm g,s}=\delta_{\rm g,s}/\delta_{\rm tot}$, where $\delta_{\rm tot}$ is the underlying total matter density contrast. The functions $b_{\rm g,s}$ are in general arbitrary functions of space and time.

In linear theory, the continuity equation  for a tracer $T={\rm g,s}$ gives
\begin{equation}
    v_T = i \frac{H\mu}{(1+z)k}\beta_{T}\delta_{T}\label{eq:conseq}
\end{equation}
where $\beta_{T}=f/b_{T}$ and $\mu \equiv \boldsymbol{\hat{r}} \cdot \boldsymbol{\hat{k}}$, where $\boldsymbol{\hat{r}}$ is the line of sight. At sub-horizon scales, this expression is also valid for most modified gravity models because they modify Einstein's equations, but not the matter conservation equations. There are also generalized models in which the conservation equations themselves are modified due to a direct interaction between matter and dark energy, so that  Eq.~\eqref{eq:conseq} has additional terms, but then typically the effect  becomes negligible at sub-horizon scales unless the interaction coupling is much larger than gravity~\citep[see e.g.][for a general energy and momentum interaction]{RyotaroTsujikawa2020a}.
Moreover, since $f$ is included in $\beta$, and we consider different $\beta$'s for different species, our formalism can also accommodate non-universal growth rates $f$, e.g. when a violation of the equivalence principle is envisaged.

We assume number densities $n_{\rm g,s}$, volumes $V_{\rm g,s}$, and a fraction $\phi=\bar n_{\rm gs}/\bar n_{\rm s}$ of supernovae hosted by galaxies in the survey, where $\bar n_{\rm gs}$ is the density of supernovae-hosting galaxies. For the cross-correlation of different tracers, we assume a common volume $V_{sg}$. If we consider the three random variables $x_{a}=\{\sqrt{V_{\rm s}}\delta_{\rm s},\sqrt{V_{\rm s}}v_{\rm s},\sqrt{V_{\rm g}}\delta_{\rm g}\}$, there are $n(n+1)/2=6$ possible 2-point correlations. Measuring  all six together forms the basis for the $6\times2$pt method proposed here. Following~\cite{Burkey:2003rk} and~\cite{Garcia:2019ita} and ignoring for now the Alcock-Paczynski (henceforth AP) corrections~\citep{Alcock:1979mp}, we can write the six observed power spectra as
\begin{align}
     \!P_{\rm gg}(k,\mu,z) &= \big[1+ \beta_{\rm g} \mu^{2}\big]^2 \,b_{\rm g}^{2} \,S_{\rm g}^2\, D_+^2 P_{\textrm{mm}}(k) + \frac{1}{n_{\rm g}}, \label{eq:pgg} \\
     \!P_{\rm ss}(k,\mu,z) &= \big[1+ \beta_{\rm s} \mu^{2}\big]^2 \,b_{\rm s}^{2}\,S_{\rm s}^2 \, D_+^2 P_{\textrm{mm}}(k) + \frac{1}{n_{\rm s}}, \label{eq:pss} \\
     \!P_{\rm gs}(k,\mu,z) &= \big[1+ \beta_{\rm g} \mu^{2}\big]\big[1+ \beta_{\rm s} \mu^{2}\big] \,b_{\rm g} \,b_{\rm s}\,S_{\rm g} \,S_{\rm s} \, D_+^2 P_{\textrm{mm}}(k) \nonumber\\
     & \quad\; + \frac{n_{\rm gs}}{n_{\rm g}n_{\rm s}}, \label{eq:pgs} \\
     \!P_{\rm gv}(k,\mu,z) &= \frac{H\mu}{k(1+z)} \!\big[1 + \beta_{\rm g}\mu^{2}\big] b_{\rm g}\, S_{\rm g}\, S_{\rm v}  \,f  D_+^2 P_{\textrm{mm}}(k), \label{eq:pgv} \\
     \!P_{\rm sv}(k,\mu,z) &= \frac{H\mu}{k(1+z)} \!\big[1 + \beta_{\rm s}\mu^{2}\big] b_{\rm s}\, S_{\rm s}\, S_{\rm v} \, f  D_+^2 P_{\textrm{mm}}(k), \label{eq:psv} \\
     \!P_{\rm vv}(k,\mu, z) &= \left[\frac{H\mu}{k(1+z)}\right]^2 S_{\rm v}^2 \,f^{2} \,D_+^2 P_{\textrm{mm}}(k)+ \frac{\sigma^2_{v, {\rm eff}}}{n_{\rm s}},
    \label{eq:pvv}
\end{align}
where $\beta_i \equiv f/b_i$,  $\mu \equiv \hat{k} \cdot \hat{r}$, $S_{\rm g,v,s}$  are damping terms  and $P_{\textrm{mm}}$ is the matter power spectrum at $z=0$. The cross-spectrum shot-noise term ${n_{\rm gs}}/{n_{\rm g}n_{\rm s}} \equiv \phi/n_{\rm g}$ in Eq. (\ref{eq:pgs}) is derived in Appendix~\ref{app:cross}.

Since one does not know \emph{a priori} the exact cosmological model, quantities such as $k$, $\mu$, and volumes  must be computed assuming a reference model. The difference between this reference and the true cosmology gives rise to AP corrections in these quantities. To wit, $\mu=\mu_{r}H/(H_{r}\alpha)$ and $k$ as $k=\alpha k_{r}$, where~\citep{2000ApJ...528...30M,Amendola:2004be}
\begin{equation}
    \alpha \,=\, \frac{H}{H_{r}} \frac{\sqrt{\mu_{r}^{2}(\eta^{2}-1)+1}}{\eta}
    \label{eq:alpha}
\end{equation}
and
\begin{equation}
    \eta \,\equiv\, \frac{HD_L}{H_{r}D_{L,r}} = \frac{HD_A}{H_{r}D_{A,r}} \,. \label{eq:eta}
\end{equation}
Both $H_r$ (the Hubble parameter) and  $D_r$ (the luminosity $L$ or angular diameter $A$ distance) are evaluated in the chosen reference ($r$) fiducial cosmology. Moreover, all observed spectra and inverse number densities in Eqs.~\eqref{eq:pgg}-\eqref{eq:pvv} get multiplied by a volume-correcting factor $\Upsilon$~\citep{1996MNRAS.282..877B,Seo:2003pu}, so that $P_{\rm xy,\,obs} \rightarrow{\Upsilon P_{\rm xy}}$, where
\begin{equation}
     \Upsilon = \frac{H D_{L,r}^2}{H_{r}D_L^2} \,.
\end{equation}
This means that in the FM these quantities also get derived with respect to whatever parameters one decides to employ. We write down all relevant derivatives of the AP terms in the Appendix~\ref{app:deriv}.

In the model-independent case, the overall factor $\Upsilon$ can be absorbed into the   power spectrum $P_{mm}$. Since this quantity is marginalized over, there is no influence of $\Upsilon$ on the model-independent forecasts: this is another consequence of the   model-independent approach being very conservative.

The non-linear smoothing factors $S_{\rm v,g,s}$, important only at small scales, have been previously approximated empirically as \citep{2014MNRAS.445.4267K,Howlett:2017asw}
\begin{align}
    S_{\rm v}=\sin(k\sigma_{\rm v})/(k\sigma_{\rm v}),\quad S_{\rm g,s}=\left[1+\frac{1}{2}(k\mu\sigma_{\rm g,s})^{2}\right]^{-1/2}\!,\label{eq:old-s}
\end{align}
with $\sigma_{\rm g}= 4.24 \,{\rm Mpc}/h$ and $\sigma_{\rm v} = 13\,{\rm Mpc}/h$. Here we adopt the unified choice
\begin{equation}
    S_{\rm v,g,s}=\exp\left[-\frac{1}{4}(k\mu\sigma_{\rm v,g,s})^{2}\right],
\end{equation}
which has the same small-$k$ limit as the expressions in (\ref{eq:old-s}). The value used in~\cite{2014MNRAS.445.4267K}
corresponds to $\sigma_{\rm v}\simeq 11$ Mpc$/h$. A more recent study by~\cite{Dam:2021fff} uses $\sigma_{\rm v}=8.5$ Mpc$/h$. We set as our fiducial values $\sigma_{\rm g}=\sigma_{\rm s} = 4.24 \;{\rm Mpc}/h$ and $\sigma_{\rm v}= 8.5$ Mpc$/h$. After testing with the methodology described below we conclude that the choice between these different models or values for $\sigma_{\rm v}$ has very little impact in our forecasts.

As can be seen, since they do not depend on galaxy bias~\citep{Zheng:2014vla}, a measurement of the peculiar velocity spectra can help considerably in breaking the degeneracy between the cosmological parameters and the nuisance bias functions. This degeneracy is the main confounding factor in measurements of the Redshift Space Distortions \citep[RSD,][]{Kaiser1987} and thus in probing the growth of structure. Previous similar works considered only one tracer of density, employing a $3\times2$pt method. If galaxies are used as density tracers and SN as velocity tracers \citep[as in][]{Howlett:2017asw} we will refer to this approach as $3\times2 \;g$--$s$. If SN trace density instead of galaxies \citep[as in][]{Amendola:2019lvy}, we will dub it $3\times2 \;s$--$s$. The full approach pursued here is similarly referred to as $6\times2 \;g$--$s$--$s$. Henceforth in this paper whenever we mention $6\times 2$pt a $6\times2 \;g$--$s$--$s$ is implied. Combining two tracers of density and one of velocity is essentially a mixture of multi-tracer and peculiar velocity approaches. The second density tracer can further help constrain the bias parameters and thus to break the degeneracy with cosmology. As is well known from multi-tracer studies~\citep{Seljak:2009,McDonald:2009,Abramo:2012,Abramo:2013,Abramo:2019ejj} the multi-tracer technique is optimal when the different tracers have very different bias, and produces no enhancement if both tracers have the very same biases. One could of course extend this to include other velocity tracers such as gravitational waves and employ, for instance, a $10\times2 \;g$--$s$--$s$--$gw$ method. Another possibility is to include also lensing spectra.

The $6\times2$pt covariance matrix is finally simply given by
\begin{equation}\label{eq:cov}
    \mathbf{C} =  \left(\begin{array}{ccc}
        \!P_{\rm gg} &  \!P_{\rm gs}  & \!P_{\rm gv} \\
        \!P_{\rm gs} &  \!P_{\rm ss}  & \!P_{\rm sv} \\
        \!P_{\rm gv} &  \!P_{\rm sv}  & \!P_{\rm vv}
    \end{array}\right).
\end{equation}
Several interesting limits can be worked out. For $n_{\rm g} \to 0$ we recover the $3\times2$pt $s$--$s$ method of~\cite{Amendola:2019lvy}. The $3\times2$pt $g$--$s$ method of~\cite{Howlett:2017asw} is instead recovered in the limit $\sigma_{\rm s}\to\infty$. The traditional full-shape density power spectrum using only galaxies ($1\times2$pt) is obtained when both $\sigma_{\rm s}$ and $\sigma_{\rm v}$ are very large. Finally, the constraints from SN velocities alone, which were discussed by~\cite{Castro:2015rrx} and~\cite{Garcia:2019ita}, are obtained when both $\sigma_{\rm g}$ and $\sigma_{\rm s}$ diverge.\footnote{
Numerically we simply drop the corresponding row(s) and column(s) to retrieve the $3\times2$pt and $1\times2$pt cases.}
With $\sigma_{\rm v}\to\infty$ we neglect the contribution of the SN velocities and reduce to a simple multi-tracer method of SN and galaxies. When one neglects the velocity information, only the Alcock-Paczynski effect in the clustering power spectra gives information on $H$. We show that inclusion of the velocity field improves the constraints substantially.

We can now simplify the problem by assuming all tracers are distributed homogeneously in the same volume $V_{\rm s}=V_{\rm g}=V_{\rm gs}=V$.\footnote{Note that we could still straightforwardly add regions or $z$-bins in which we have only galaxies or only SN, treating them separately with a $3\times2$pt or $1\times2$pt approach.} In this case we write the FM for a set of parameters $\theta_{\alpha}$, in a survey of volume $V$, as~\citep{Tegmark:1997rp,Abramo:2019ejj}
\begin{equation}
    F_{\alpha\beta} \,=\, \frac{1}{(2 \pi)^3} 2\pi k^{2}\Delta_{k}V\bar{F}_{\alpha\beta} \,=\, VV_{k}\bar{F}_{\alpha\beta}\,,
\end{equation}
where $V_{k}= (2\pi)^{-3}2\pi k^{2}\Delta_{k}$ is the volume of the Fourier space after integrating over the azimuthal angle but not over the polar angle (i.e., the volume of a spherical Fourier space shell of width $\Delta_{k}$ would be given by $\int \dd \mu \, V_k$). In this expression $\bar{F}$ is the FM per unit phase-space volume $V V_k$ integrated over~$\mu$,
\begin{equation}
    \bar{F}_{\alpha\beta} = \frac{1}{2}\int_{-1}^{+1} \dd\mu \, \frac{\partial C_{ab}}{\partial\theta_{\alpha}}C_{ad}^{-1}\frac{\partial C_{cd}}{\partial\theta_{\beta}}C_{bc}^{-1} \,,
\end{equation}
where the integrand is evaluated at the fiducial value.

Denoting with $V(z)$ the volume of the $z$-shell, the $k$-cells are chosen with size $\Delta_{k}=2\pi/V(z)^{1/3}$ between $k_{\rm min}(z)$ and $k_{\rm max}$. Following~\cite{Garcia:2019ita}, we take $k_{\rm min}=2\pi/V(z)^{1/3}$,  while $k_{\rm max} = 0.1~h/$Mpc ensures we remain in the linear regime. We shall briefly explore using lower or higher values of $k_{\rm max}$, to wit $0.05$ and $0.15 \,h/$Mpc.

For the model-independent approach, it was shown by~\citet{Amendola:2019lvy} that $H_{0}D$ is typically known with relatively much higher accuracy from SNe magnitudes alone, and therefore a constraint on $\eta$ becomes entirely equivalent to a constraint on $E(z)=H/H_{0}$.

It is important to notice that the bias $b_{\rm g,s}$ are not expected to coincide. The bias in a population of galaxies depends on the specific mix of luminosities selected at a particular redshift, and the bias-luminosity relationship is relatively well-understood. The number density of SN, on the other hand, depends in complicated ways on the characteristics of their local environment and of the host galaxy. For our purposes the key fact is that, in a multi-tracer approach, the constraints on cosmological parameters are stronger when the biases of each population are different, so we expect the same trend here, i.e. stronger constraints when $b_{\rm g}/b_{\rm s}\not=1$.

\section{Surveys and fiducials}\label{sec:surv}

We will forecast results for two future surveys. One, dubbed simply ``Conservative'', will assume a galaxy coverage as given by the 4MOST survey and SN detected by LSST during a 5-year survey and appropriately followed-up in order to allow classification, assuming a constant completeness of $15\%$ in the redshift range $0<z<0.4$ and a joint area of 7500 deg${}^2$. The second survey, dubbed simply ``Aggressive'', uses galaxies and SN assuming instead a constant completeness of $30\%$ in the redshift range $0<z<0.7$ in an area of 18000 deg${}^2$, equivalent to the whole LSST area. The Aggressive case will definitely pose a challenge due to the large amount of SN that would need to be properly classified, which means we would need to rely on photometrically classified SN~\citep[see, e.g.][]{Lochner:2016hbn}.
\cite{VargasdosSantos:2019ovq} has nevertheless discussed how a Bias-Variance tradeoff approach might be employed in order to account for non-Ia contaminants, and found that current Machine Learning techniques of photometric classification resulted in an effective SN completeness between 33\% and 75\%. This means in practice that unless the photometric classification methods are improved, we would have only effectively this fraction of SN out of the total observed ones.

The 4MOST Consortium Survey 8: Cosmology Redshift Survey (CRS) \citep[][]{Richard2019} is a $7500 \,\rm deg^2$ galaxy and quasar survey designed to overlap fully with LSST. It will observe three galaxy populations: bright low-redshift galaxies (BGs), luminous red galaxies (LRGs) and emission line galaxies (ELGs). It will also observe two quasar populations which we will not consider here. Definitive predictions for the galaxy bias values of these populations are yet to be established. Previous results for BOSS, eBOSS and DESI show that common changes in the target selection strategy—such as color range cuts and photometric filter choice—can have an impact on the overall bias normalization at the 10--20\% level~\citep{Zhou2021}. We choose to follow specifications from the DESI Collaboration BGS and LRG samples \citep[][]{desicollaboration2016}, which are sufficiently close to the expected 4MOST BG and LRG samples for the purposes of our analysis. Therefore, we combine the samples' redshift distributions, and take the overall galaxy bias $b_{\rm g}$ to be $1.34/D_+(z)$ for $z\le0.3$ (mostly BGs) and $1.7/D_+(z)$ for $z>0.3$ (mostly LRGs).

\begin{table}
    \footnotesize
    \setlength{\tabcolsep}{3.7pt}
    \begin{tabular}{ ccc | cc | cc}
    \hline
    & \multicolumn{3}{c}{\rm{Conservative}} & \multicolumn{3}{|c}{\rm{Aggressive}} \\
    $z_{\rm bin}$ & V & $10^{3}\cdot n_{\rm s}$ &  $b_{\rm g}$ & V & $10^{3}\cdot n_{\rm s}$ & $b_{\rm g}$ \\
      & $(\text{Gpc}/h)^{3}$ & $(h/\text{Mpc})$ &  & $(\text{Gpc}/h)^{3}$ & $(h/\text{Mpc})^3$ &  \\
    \hline
     0.05 & 0.019 & 0.048 & 1.38 & 0.046 & 0.096  & 1.38    \\
     0.15 & 0.123 & 0.052 & 1.45 & 0.296 & 0.105  & 1.45     \\
     0.25 & 0.303 & 0.057 & 1.53 & 0.727 & 0.114  & 1.53    \\
     0.35 & 0.531 & 0.061 & 2.04 & 1.27  & 0.122  & 2.04    \\
     0.45 &  --   &  --   &  --  & 1.88  & 0.131  & 2.15    \\
     0.55 &  --   &  --   &  --  & 2.51  & 0.139  & 2.26    \\
     0.65 &  --   &  --   &  --  & 3.13  & 0.148  & 2.37    \\
     \hline
    \end{tabular}
    \caption{\label{tab:sn-surveys}  Survey specifications for  both scenarios forecast here. In both cases we assume supernova are detected by LSST with spectra obtained in follow-up surveys and that $n_{\rm g} = 10 n_{\rm s}$. For the Conservative (Aggressive) forecast we assume a 15\% (30\%) completeness of SNe in all redshifts covering a total area of 7500 deg$^{2}$ (18000 deg$^{2}$) in the range $0<z<0.4$ ($0<z<0.7$). The supernova bias is assumed to be either $1.0/D_+(z)$ or $1.5/D_+(z)$; the galaxy bias is assumed to be $1.34/D_+(z)$ for $z\le0.3$ (mostly BGs), $1.7/D_+(z)$ for $z>0.3$ (mostly LRGs).  The $z$ bins have $\Delta z=0.1$ and are centred on $z_{\rm bin}$.
    }
\end{table}

For the number density of galaxies we will assume, for simplicity,  a baseline  such that  $n_{\rm g} = 10 \,n_{\rm s}$. This is a deliberate, conservative choice, which should be easily achievable with spectroscopic surveys like 4MOST and DESI. For this reason we do not consider here photometric redshift errors. For the number density of SN, we use the SN rate $\mathtt{r}_{Ia} = 2.1 \cdot 10^{-5} (1+z)^{1.95}$/(yr Mpc${}^3$),  which is a good fit to~\cite{Cappellaro:2015}. The observed number densities in units of (Mpc$/h){}^3$ for a 5-year survey are then given by
\begin{equation}
    n_s = \frac{5}{h^3} \, 2.1 \cdot 10^{-5} (1+z)^{0.95}  \, \mathcal{C}_{Ia}\,,
\end{equation}
where we divided $\mathtt{r}_{Ia}$ by $1+z$ to transform the rest-frame rates into observed rates and where $\mathcal{C}_{Ia}$ is the assumed supernova completeness of the survey. This means that the total number of SN is around 57 thousand in the Conservative case ($0<z<0.4$) and 1.3 million in the Aggressive case ($0<z<0.7$). We will also briefly explore a higher galaxy density scenario in which $n_{\rm g} = 100 \,n_{\rm s}$. This broad range includes the expected 4MOST CRS densities, and the use of simple ratios $r\equiv n_{\rm g} / n_{\rm s}$ between tracer densities allow for a clearer presentation of the benefits of a $6\times2$pt approach. We will also show that the gain saturates for $r\ge10$. As pointed out above, the assumed fraction $\phi$ of supernovae inside galaxies included in the survey has negligible impact on our results because we always consider $r \gg 1$: the uncertainties forecast change by less than $1\%$ whether $\phi$ is 0~or~1.

As for the supernova bias $b_s$, little is currently known. A recent analysis \citep{MukherjeeWandelt2018} takes results from \citet{Carlberg2008}, whose measurement of the angular correlation function is specific to the SNLS 3-year survey \citep{Guy2010}, with low SNR, and in the redshift range $0.2 < z < 0.9$. There is no specific reason that the bias value thus obtained should be equal or very similar for more general populations. Faced with this uncertainty, we choose to assume two bias models for our SN population, $1.0/D_+(z)$ or $1.5/D_+(z)$, corresponding to a low and high bias cases, respectively.

We note that in realistic cases the fraction of SNe that populate a known galaxy survey can imply a relation between $\phi$, $b_g$ and $b_s$, depending on how this subsample of SNe is assumed to populate the galaxy parent sample (i.e. randomly or, if not, how). Given the lack of any current information on this issue, we choose to treat $\phi$ as independent from both $b_g$ and $b_s$ in order to keep the analysis more transparent.

The fiducial power spectrum $P(k)$ is obtained with CAMB\footnote{\url{https://camb.info/}}~\citep{Lewis:1999bs}  from $\Lambda$CDM with $\Omega_{m0}=0.3$, $h=0.7$, $\sigma_8 = 0.83$ and the rest of the parameters set to Planck 2018 values~\citep{Planck:2018vyg}. We also include the non-linear Halofit corrections as implemented in CAMB~\citep{Lewis:1999bs,Takahashi:2012em}, although this has a minor effect since we cut at $k_{\rm max}=0.1$ Mpc$/h$. For the growth-rate index we set a fiducial $\gamma = 0.545$. We assume throughout a constant SN intrinsic scatter $\sigma_{\mathrm{int}}=0.13$~mag and non-linear velocity scatter $\sigma_{v{\rm ,nonlin}}=300$ km/s. These are similar values to those used in recent SN compilations such as the Pantheon+~\cite{Peterson:2021hel} (which used 0.14 mag and 250 km/s) and JLA~\cite{Betoule:2014frx} (which used values between 0.08-0.12 mag and 150 km/s, in addition to a lensing scatter).\footnote{Interestingly, near infrared light-curve measurements have been shown to exhibit smaller dispersions when compared to optical ones by as much as 35\%~\citep{Avelino2019}.}
The details for both Conservative and Aggressive surveys here considered are summarized in Table~\ref{tab:sn-surveys}.  We make use throughout of broad redshift bins with width $\Delta z = 0.1$. We assume that the correlation among these bins can be neglected due to their relatively large size.

In the model-dependent approach we will retain as free cosmological parameters $\sigma_8$, $\gamma$, $\Omega_{m0}$, $\Omega_{k0}$ and $h$, while other parameters which are less relevant (such as the primordial power spectrum slope $n_{\rm s}$, the optical depth $\tau$, and the baryonic fraction $\Omega_{b0}$) will be kept fixed for simplicity at the respective Planck best fit values. For $\sigma_8$, $\gamma$ and $h$ we use flat uninformative priors, while for both $\Omega_{m0}$ and $\Omega_{k0}$ we use very broad Gaussian priors with $0.5$ uncertainties. These two last quantities are both background quantities well constrained by many different observables, and tighter priors would easily be justifiable. We nevertheless opt to show results using broad priors so that the final (loose) constraints on $\Omega_{k0}$ become visible and because the priors on the background quantities have little effect on $\sigma_8$ and $\gamma$.  All nuisance parameters  will be left free to vary with  weak priors, which we set to be Gaussians with $50\%$ relative uncertainties. We will employ three global nuisance parameters to account for the non-linear RSD ($\sigma_{\rm g},\,\sigma_{\rm s},\,\sigma_{\rm v}$) and two bias nuisance parameters \emph{in each redshift bin} ($b_{\rm g}^{z_i}$ and $b_{\rm s}^{z_i}$). This means that we have a total of 11 (17) nuisance parameters for the Conservative (Aggressive) forecasts. The final parameter vector is thus:
\begin{equation}
    \boldsymbol{\theta} = \{ \sigma_8, \gamma, \Omega_{m0}, \Omega_{k0}, h, b_{\rm g}^{z_i}, b_{\rm s}^{z_i} , \sigma_{\rm g}, \sigma_{\rm s}, \sigma_{\rm v}\}\,.
\end{equation}

\begin{figure*}
    \includegraphics[width=2.08\columnwidth]{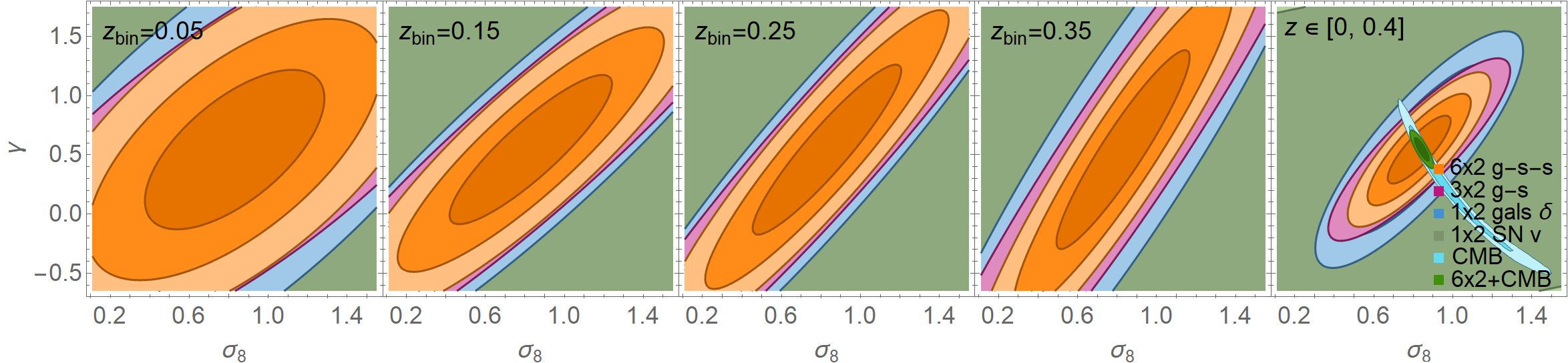}
    \caption{
    Forecasts for the Conservative case (see Table~\ref{tab:sn-surveys}), assuming $k_{\rm max} = 0.1 h/$Mpc, $b_{\rm s} = 1.0/D(z)$ and redshift bins with $\Delta z = 0.1$. The last panel shows the aggregate constraints considering all redshift bins. The 1, 2 and $3\sigma$ contours are as follows: orange for the full $6\times2$pt method; magenta for the $3\times2$pt $g$-$s$ (no tracing of density with SNe); blue are the traditional $1\times2$pt of galaxies (the full-shape power spectrum, using only density tracers); green is using only velocity tracers (with SNe). Only in the last panel, and only 1 and 2$\sigma$: cyan are the current CMB constraints~\citep[from][]{Mantz:2014paa} and dark green the combination of CMB and \six.  The FoM of \six\ is 38\% (125\%) higher than \three\ ($1\times2$pt of galaxies).
    \label{fig:gamma-sigma8-cons}}
\end{figure*}

In the model-independent case, we take as free model-independent parameters the following set:
\begin{equation}
    \boldsymbol{\theta}=\{P^{k_i,z_j}, \beta_{\rm g}^{k_i,z_j},\beta_{\rm s}^{k_i,z_j},\eta^{z_j},\sigma_{\rm g},\sigma_{\rm s},\sigma_{\rm v}\}\,,\label{eq:parmodind}
\end{equation}
where $P=n_{\rm s}b_{\rm s}^{2}D_+^2 P_{mm}$ is the supernovae signal-to-noise spectrum and $\eta$ has been defined in Eq. (\ref{eq:eta}).
The first three parameters are left free to vary for each $k,z$ cell; the fourth, $\eta$,  varies only in each $z$-shell; the last three, the $\sigma$'s, are assumed independent of both $k$ and $z$. The total number of free parameters depends therefore on how many $k,z$ cells we consider. The $3\times2$pt $g$--$s$ model-independent approach was discussed in detail in the original CSC paper~\citep{Amendola:2019lvy}. The extension of this method to a $6\times2$pt $g$--$s$--$s$ is straightforward.  In the model-independent approach, all the priors are uninformative, except for $\sigma_{\rm g,s,v}$, as detailed later on. The parameter $h$ is not included in this approach because degenerate with the other quantities.

We also investigated in both approaches how much precision is lost if one allows instead all three non-linear RSD parameters $\sigma_{\rm g,s,v}$ to vary freely in each redshift bin. As we will discuss below, the results for the Conservative case show negligible variations and only for the Aggressive survey specifications would this lead to a small degradation of the forecasts.

All our results presented here for a given parameter or parameter pair are fully marginalized over all the other parameters.

\section{Results for $\gamma$, $\sigma_8$ and bias parameters in the model-dependent approach}

\begin{figure*}
    \includegraphics[width=1.98\columnwidth]{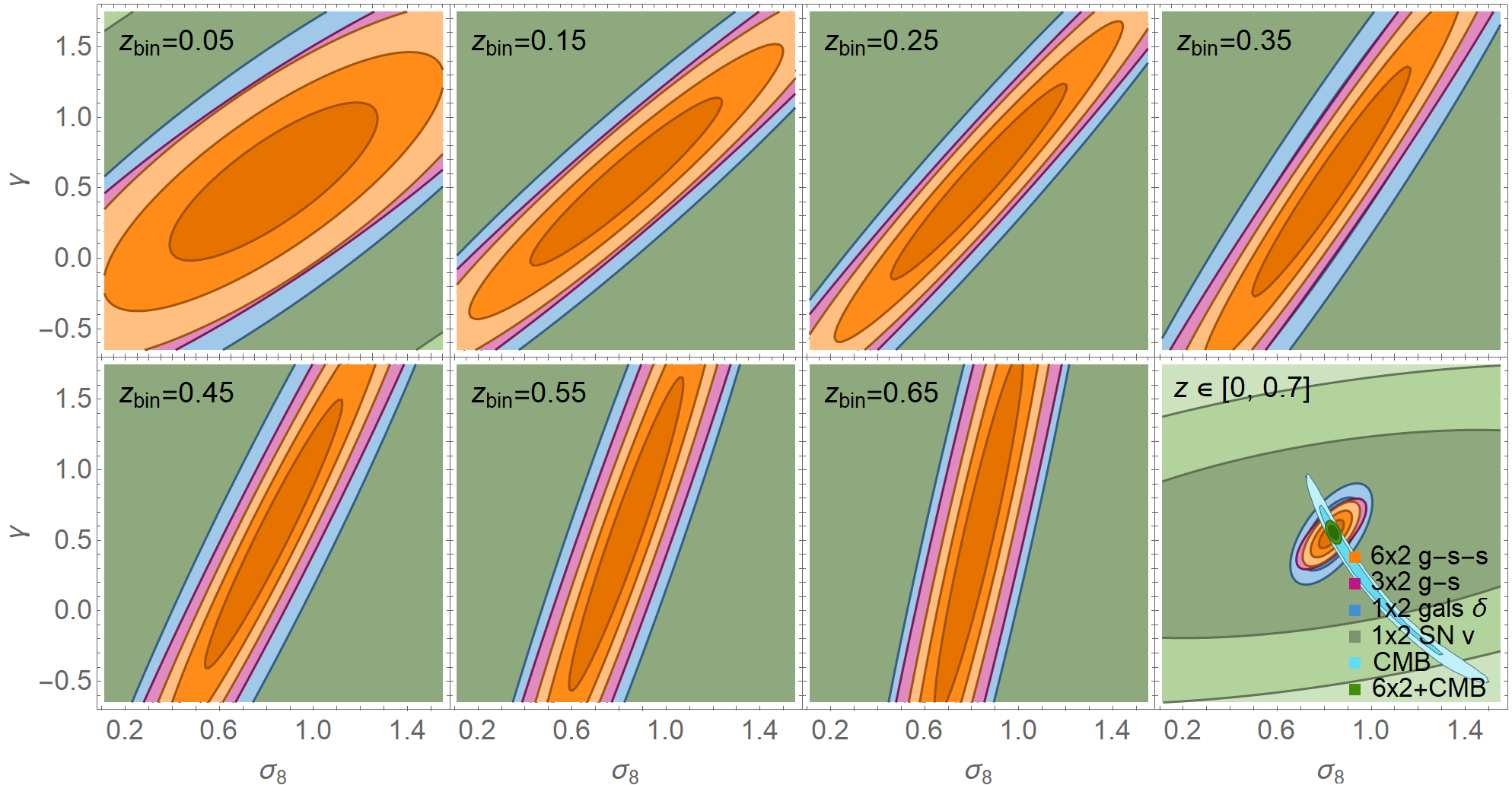}
    \caption{
    Same as Figure~\ref{fig:gamma-sigma8-cons} for the Aggressive case (which has larger area and SN completeness -- see Table~\ref{tab:sn-surveys}). The lower relative information from velocities at higher redshifts result in more  modest gains for the $3\times2$pt when compared with the ($1\times2$pt) full-shape power spectrum constrains. The FoM of $6\times2$pt is 28\% (145\%) higher than $3\times2$pt ($1\times2$pt of galaxies).
    \label{fig:gamma-sigma8-aggr}}
\end{figure*}

Here we present the forecast constraints for $\gamma$, $\sigma_8$, $\Omega_{m0}$ and the bias parameters (which are left free in each redshift bin). We compare the constraints obtained through four methods: two \one\ cases (using only galaxies as density tracers or only SN as velocity tracers), the \three\ $g\!-\!s$ case where SN velocities complement galaxy densities and the full \six\ case. Since our focus is on $\gamma$ and $\sigma_8$, we also define a figure of merit (FoM) as the square-root of the determinant of the $2\times2$ FM in these parameters after marginalising over all other ones.

Table~\ref{tab:FOM} summarises the results on the FoM in all cases here considered. To wit, the Conservative and Aggressive survey specifications, the low and high values of the galaxy number densities $n_{\rm g}$ and the low and high values of the SN bias $b_{\rm s}$. The \six\ case outperforms the \three\ approach by between 10 and 37\%, depending on the assumptions. Gains are higher for low SN bias. The reason is that $b_{\rm s}$ becomes more distinct from $b_{\rm g}$, and the \six\ method becomes degenerate with \three\ if $b_{\rm s}=b_{\rm g}$, as in any multi-tracer method. Gains are a bit smaller when $n_{\rm g}$ is much larger than $n_{\rm s}$. Note that in all cases, both \three\ and \six\ methods yield improvements of a factor of two or larger when compared to the traditional \one\ approach which ignores the velocity spectra.

In the rest of this section for simplicity we focus on the case $b_{\rm s} = 1 / D_+$ and $n_{\rm g} = 10 n_{\rm s}$. Figure~\ref{fig:gamma-sigma8-cons} and Figure~\ref{fig:gamma-sigma8-aggr} show the forecast 1, 2 and $3\sigma$ contours on $\gamma$ vs $\sigma_8$ for four different combinations of spectra for the Conservative and Aggressive surveys, respectively. We depict the traditional \one\ approach, the case of using only SN velocities (i.e. only $P_{vv}$), as well as the \three\ $g$--$s$ and \six\ $g$--$s$--$s$ methods. We also illustrate both the constraints in each $z$-bin and the combined constraints in all bins considered. In the lowest redshift bins the constraints using only $P_{vv}$ are tighter than in higher redshifts as a result of the larger relative contribution of peculiar velocities in nearby galaxies where the Hubble flow is smaller. Nevertheless by itself $P_{vv}$ carries little cosmological information. But combining it with density spectra results in substantial improvements even on higher redshifts, as can be seen in the gains obtained using \three. So velocities add important information in all redshifts. Finally, $P_{vv}$ is the most sensitive to the assumed priors; using tight priors on the background quantities yield large improvements in the $P_{vv}$-only constraints, and we recover results consistent with~\cite{Garcia:2019ita}.

\begin{table}
    \footnotesize
    \setlength{\tabcolsep}{2.7pt}
    \begin{tabular}{ c c c c c c c}
    \hline
    & \multicolumn{3}{c}{\rm{Conservative}} & \multicolumn{3}{|c}{\rm{Aggressive}} \\
    case & only $g$ & $3\times2$pt & $6\times2$pt & only $g$ & $3\times2$pt & $6\times2$pt  \\
    \hline
    \parbox[c][1.1cm]{1.8cm}{$\frac{n_{\rm g}}{n_{\rm s}}=10$\\$b_{\rm s}=\frac{1.0}{D_+(z)}$}
    & 40 & 66 & 91 & 240 & 460 & 590   \\
    \hline
    \parbox[c][1.1cm]{1.8cm}{$\frac{n_{\rm g}}{n_{\rm s}}=100$\\$b_{\rm s}=\frac{1.0}{D_+(z)}$}
    & 43 & 69 & 91 & 250 & 480 & 600   \\
    \hline
    \parbox[c][1.1cm]{1.8cm}{$\frac{n_{\rm g}}{n_{\rm s}}=10$\\$b_{\rm s}=\frac{1.5}{D_+(z)}$}
    & 40 & 66 & 86 & 240 & 460 & 520   \\
    \hline
    \end{tabular}
    \caption{\label{tab:FOM}  Figure of merit (FoM) in the $\sigma_8 \times \gamma$ plane marginalizing over all other variables and nuisance parameters for the different forecast assumptions. As can be seen, there is little extra information for $n_{\rm g} > 10 n_{\rm s}$. Precision in the \six\ approach also approaches the one of \three\ when $b_{\rm s}$ gets closer to $b_{\rm g}$.
    }
\end{table}

In the last panels depicting the combined constraints using all redshifts we also plot the current CMB constraints from~\cite{Mantz:2014paa}, which used Planck 2013 temperature data combined with WMAP polarisation data and high-multipole ACT~\citep{Das:2013zf} and SPT data~\citep{Story:2012wx}, assuming flatness. Note that the CMB constraints are completely orthogonal to our LSS forecasts. As can be seen, a combination of \six\ approach with CMB can yield very high-precision measurements. On the other hand it makes it harder for large systematics effects to introduce tensions and therefore be made apparent \emph{a posteriori}~\citep{March:2011rv}.

\begin{figure}
    \includegraphics[width=1.0\columnwidth]{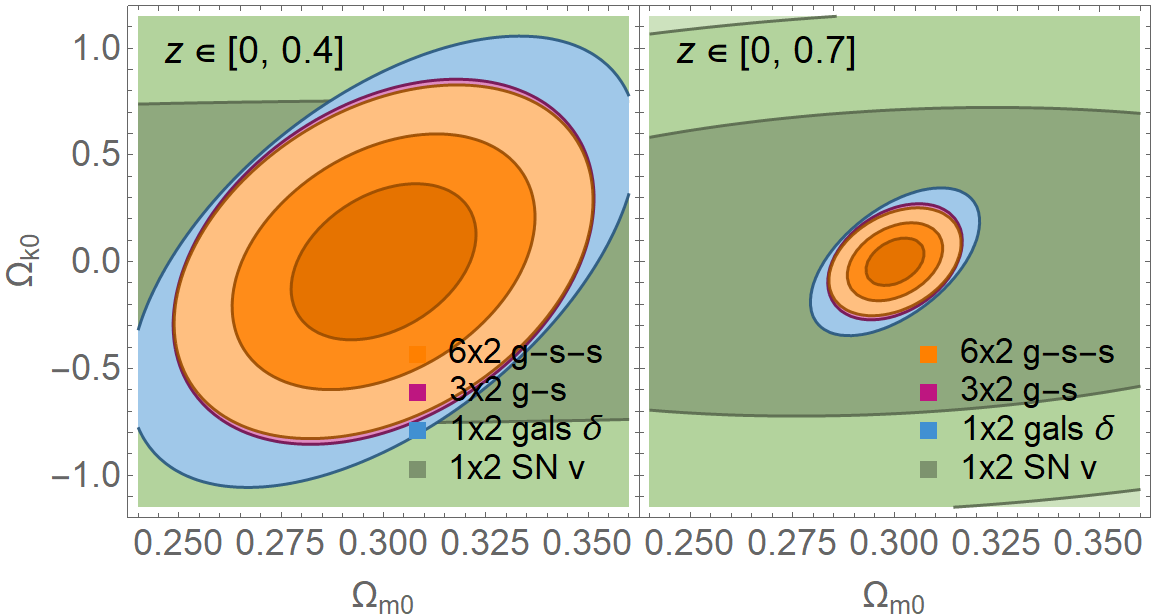}
    \includegraphics[width=1.0\columnwidth]{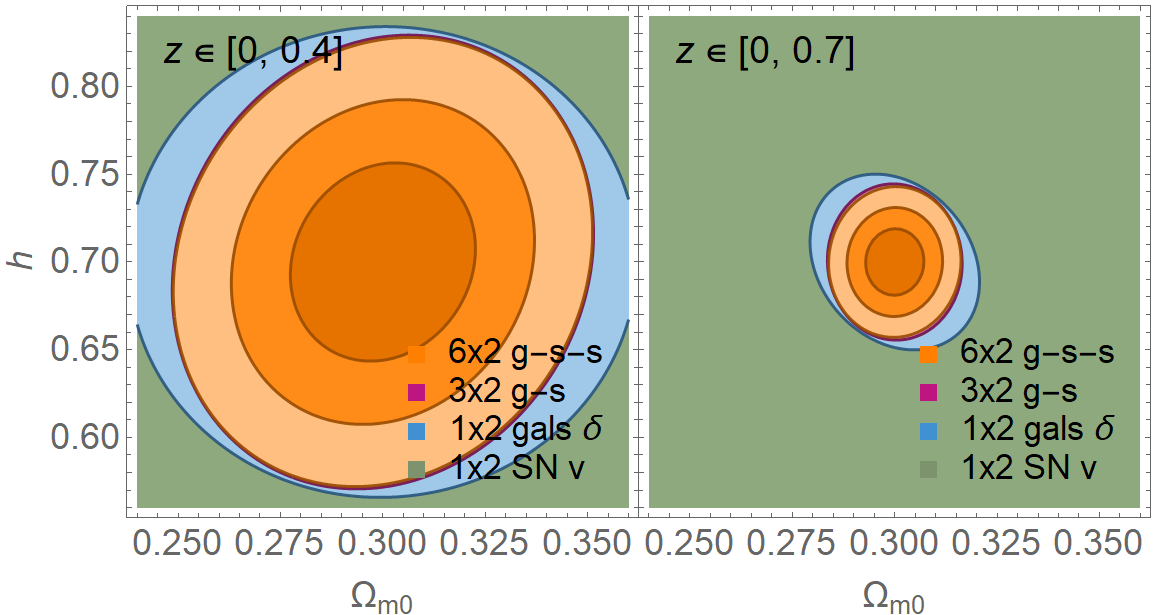}
    \caption{Similar to Figures~\ref{fig:gamma-sigma8-cons}  and~\ref{fig:gamma-sigma8-aggr} for the constraints on $\Omega_{k0}$ vs. $\Omega_{m0}$ and on $h$ vs. $\Omega_{m0}$ , considering all redshift bins together. The left (right) panels depict the Conservative (Aggressive) forecasts. \label{fig:sigma8-Om0}}
\end{figure}

Figure~\ref{fig:sigma8-Om0} shows both the  $\Omega_{k0}$ vs. $\Omega_{m0}$ and the $h$ vs. $\Omega_{m0}$ constraints using all redshifts. The \three\ and \six\ approaches yield almost identical results, but both show small gains in $\Omega_{m0}$, $\Omega_{k0}$ and $h$. The density parameters $\Omega_{k0}$ and $\Omega_{m0}$ show some correlation among themselves but all three background parameters have little correlation with $\sigma_8$ and $\gamma$.

The final, fully marginalized forecast uncertainties in each parameter is shown in Table~\ref{tab:1Derrors}. In this table we focus on the \six\ results and explore different cases: (i) in which flatness is assumed; (ii) in which different values of $k_{\rm max}$ are used; (iii) in which the AP corrections are neglected. With the default assumptions we see in the Conservative case that $\sigma_8$ and $\gamma$ are constrained to within $12\%$ and $35\%$ respectively, and that the Hubble constant can be measured with $5\%$ uncertainty.  We also show CMB constraints and the joint \six\ + CMB results. As can be seen, the separate constraints are somewhat comparable, but the joint constraints offer a huge improvement due to the orthogonality of the degeneracies. The combined Conservative + CMB results yield improvements of factors of 5 in \emph{each} cosmological parameters compared to CMB alone. This allows for instance final $0.4\%$ constraints on curvature, a tighter constraint than even Planck data including CMB lensing. We note however that the CMB results are in fact an approximation: the constraints on the parameters \{$\Omega_{m0}$,$\,\Omega_{k0}$,$\,h$\} are those obtained from the marginalized Planck 2018 TTTEEE chains~\citep{Planck:2018vyg} including curvature, while those in the parameters \{$\sigma_8$,$\,\gamma$\} are derived from the chains computed by~\cite{Mantz:2014paa} assuming flatness and using a modified version of CosmoMC \citep{Rapetti2009}. Since the chains have non-Gaussian  posteriors (for $\gamma$ it is even multimodal), for simplicity we quote the standard deviation of the posteriors instead of the exact asymmetric uncertainties.
A full Planck 2018 analysis including $\gamma$ and $\Omega_{k0}$ is not available to our knowledge. In Appendix~\ref{app:cmb-combination} we comment further on these CMB constraints.

Neglecting AP (both the $\Upsilon$ term and the $k$ and $\mu$ corrections) can bias the cosmological results, but it is not straightforward to estimate this bias with a FM methodology. In fact, the methodology for estimating systematic shifts in parameters proposed by~\cite{Huterer:2004tr} and \cite{Amara:2007as} relies on bias shifts in the covariance matrix $\mathbf{C}$ of Eq.~\eqref{eq:cov}, but we assume throughout that the reference cosmology is also the best-fit cosmology, and thus both including or not AP, $\mathbf{C}$ remains unchanged. In other words, the systematic effect of neglecting AP is proportional to the difference between the reference and true cosmologies, which is not a straightforward quantity to estimate a priori. However, we clearly notice a slight increase in the uncertainties of $\sigma_8$ and $\gamma$ (around 5\%), of both $\Omega_{m0}$ and $\Omega_{k0}$ (around 30\%) and, most of all, $h$ (90\%). These results show the importance of fully taking the AP effect into account, and how it at the same time contains extra information but is also partially degenerate with the RSD, thus introducing confounding factors.

As expected, the assumed value of $k_{\rm max}$ plays an important role on the forecasts. We first note that in principle at higher redshifts one is expected to rely on slightly higher $k_{\rm max}$  than at low redshift. Here instead we simply assume a constant value of $0.1 \, h$/Mpc and explored also the cases of $0.05$ and $0.15\, h$/Mpc. In a retrospective study~\cite{Foroozan:2021zzu} found out that previous FM forecasts match current observational results best when one assumes an effective $k_{\rm max} = 0.08 \, h$/Mpc.  Given that there is a large community effort at the moment to understand how accurately one can keep non-linearities under control and how to extract smaller scale information reliably, it is hoped that using (effective) wavenumbers larger than $0.08 \, h$/Mpc may be possible in the near future.

\begin{table}
    \centering
    \setlength{\tabcolsep}{4.0pt}
    \begin{tabular}{c c c c c c}
    \hline
    $1\sigma$ uncertainties in:  & $\sigma_8$ & $\gamma$ & $h$ & $\Omega_{m0}$ & $\Omega_{k0}$  \\
    \hline
    Conservative               & 0.10   & 0.19  & 0.037 & 0.015 & 0.24     \\
    Conservative [\three]      & 0.13   & 0.22  & 0.038 & 0.015 & 0.25     \\
    Conservative [no AP]       & 0.11   & 0.20  & 0.070 & 0.019 & 0.36     \\
    Conservative [flat]        & 0.10   & 0.19  & 0.028 & 0.014 & -        \\
    Conserv. [$k_{\rm max}=0.05$] & 0.15  & 0.28 & 0.12  & 0.031 & 0.39     \\
    Conserv. [$k_{\rm max}=0.15$] & 0.091 & 0.16 & 0.019 & 0.010 & 0.19     \\
    \hline
    Aggressive                 & 0.036  & 0.067 & 0.013 & 0.0047 & 0.074   \\
    Aggressive [\three]        & 0.043  & 0.073 & 0.013 & 0.0048 & 0.079   \\
    Aggressive [flat]          & 0.034  & 0.066 & 0.0085 & 0.0044 & -      \\
    \hline
    CMB [${}^\star$]           & 0.18 & 0.34 & 0.037 & 0.064 & 0.017    \\
    Conservative + CMB         & 0.022  & 0.058 & 0.0073& 0.010 & 0.0037   \\
    Aggressive + CMB           & 0.013  & 0.032 & 0.0045 & 0.0036 & 0.0028 \\    
    \hline
    \end{tabular}
    \caption{\label{tab:1Derrors} Fully marginalized absolute forecast uncertainties in each cosmological parameter using the \six\ method (unless otherwise stated).  Assuming flatness leads to better precision in $\gamma$ and $h$, but has almost no effect on the constraints on $\sigma_8$ and $\Omega_{m0}$. Ignoring the AP corrections (``no AP'' row) leads to slightly incorrect constraints especially in the background parameters (and to biases in the case of real data). We also show results for  $k_{\rm max} = 0.05$ and $0.15 \,h/$Mpc. CMB constraints are an approximation of Planck results [see text].
    }
\end{table}

\begin{figure}
    \centering
    \includegraphics[width=.95\columnwidth]{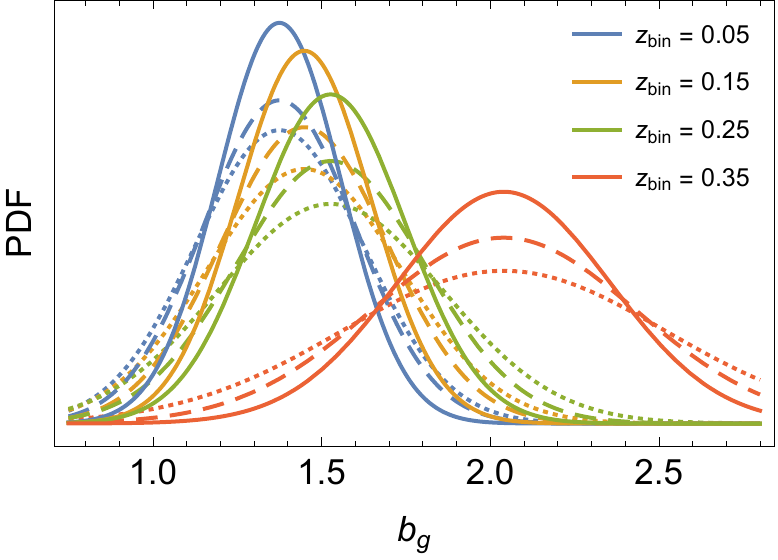}
    \caption{Forecast PDFs on the different galaxy bias parameters for the Conservative case. Solid lines: \six\ constraints, dashed lines: \three\ constraints; dotted \one\ constraints.
    \label{fig:biases}}
\end{figure}

Finally, on Figure~\ref{fig:biases} we plot the forecast PDFs on the different bias parameters, which are left free to vary in each redshift bin. The \six\ method is able to independently measure all bias parameters with better precision than \three\ and \one\ methods. This makes the \six\ approach much more robust as it is less sensitive to bias modeling.  The \six\ relative uncertainties in all bias parameters are around $15\%$ ($5\%$) for the Conservative (Aggressive) surveys, with little variation in redshift. For comparison, the \one\ constraints in $b_g$ instead are around $21\%$ ($7\%$) for the Conservative (Aggressive) cases.

\section{Results for $H(\lowercase{z})/H_0$ in the model-independent approach}

In the model-independent approach we  do not assume any parametrization. The cosmological quantities are therefore left free to vary in each $z,k$ bin (taken to be  the same as in Sect. \ref{sec:surv}). We focus on the forecasts for $H(z)/H_0$.

The only methodological difference with respect to previous work, is that in \cite{Amendola:2019lvy} we transformed $k$ and $\mu$ but not $\sigma_{\rm v,g,s}$. However, to transform a velocity into a distance one needs to use a model for $H(z)$: $\Delta r=\Delta z/H(z)$. Therefore the transformation law  should actually be $k\mu\sigma\to k_r\mu_r\sigma_r\eta H_{r}/H=k_r\mu_r\sigma_r D/D_r$ and therefore be independent of $H(z)/H$. Beside this, the approach is  the same as in~\cite{Amendola:2019lvy}, to which we refer for more details. As in that paper, we  assume that the luminosity distance of the supernovae can be estimated to high precision so we can neglect this uncertainty with respect to the one we find for $H(z)/H_0$. For this assumption to hold it is sufficient to assume $D_L$ to be a smooth function of $z$.

In this section we adopt the Aggressive  case of Table \ref{tab:sn-surveys}, with the same specifications discussed earlier, except when otherwise indicated. In particular, when varying $r=n_g/n_s$, we keep $n_s$ fixed and vary $n_g$.

\subsection{Dependence on the prior for $\sigma_{\rm v,g,s}$}

\begin{figure}
\includegraphics[width=.98\columnwidth]{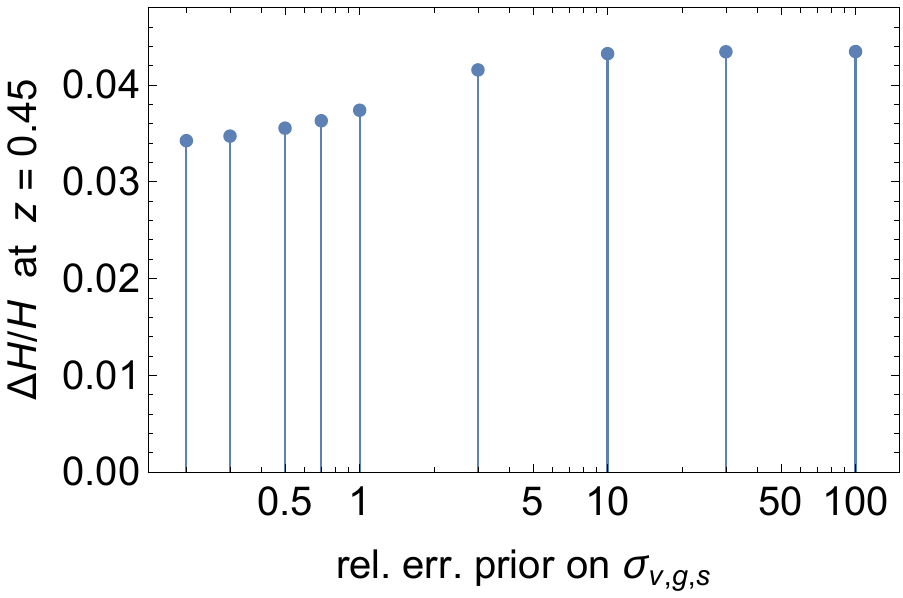}
\caption{Dependence of the relative error for $H$ on the prior of $\sigma_{\rm v,g,s}$ (in units of their relative error) .  \label{fig:dep-sigma}}
\end{figure}

We begin the analysis by showing how the constraints on $H(z)/H_0$ depend on the standard deviation of the Gaussian prior on the non-linear smoothing $\sigma_{\rm v,g,s}$ (see Figure~\ref{fig:dep-sigma}). The constraints improve by roughly only 30\% when tightening the prior from a relative standard deviation of 100\% to 20\%. Since the trend is very weak,  we choose  a conservative prior of 50\%.

\subsection{Dependence on $n_{\rm g} / n_{\rm s} $}

\begin{figure}
\includegraphics[width=.98\columnwidth]{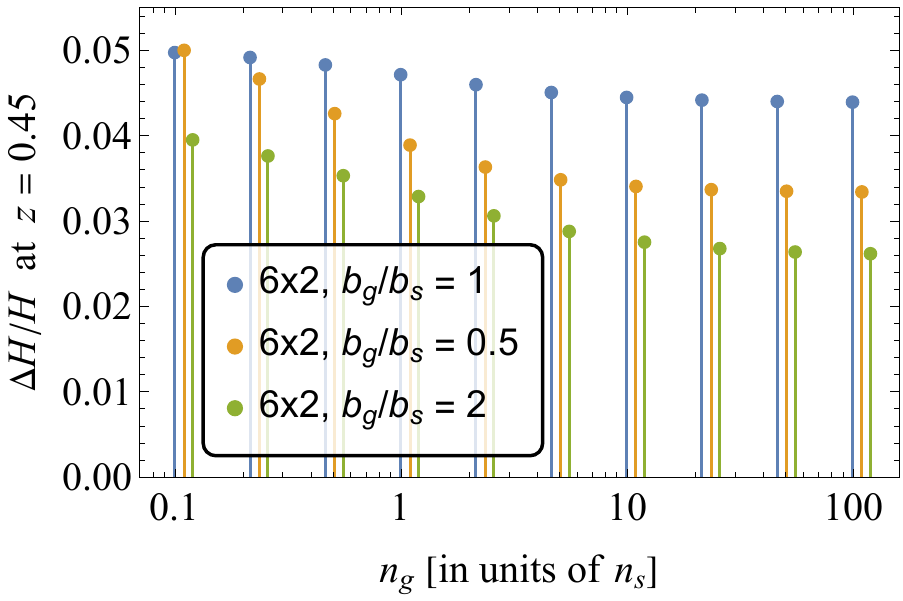}
\caption{Dependence on $n_{\rm g}/n_{\rm s}$ (keeping $n_s$ fixed) of the relative error on $H$ for  50\% prior on $\sigma_{\rm v,g,s}$ and various values of the bias ratio obtained varying $b_s$. Orange and green bars are slightly displaced for clarity. \label{fig:dep-r}}
\end{figure}

We now show how the constraints on $H$ depend on $r=n_{\rm g}/n_{\rm s}$.  Since we are interested in exploring the dependence on various assumptions, we begin by assuming equal bias for galaxies and supernovae,  $\beta_{\rm g}=\beta_{\rm s}$, with $\beta_{\rm g}$ given in Sect. \ref{sec:surv}.  For simplicity, we present the relative marginalized errors on $H$ only for a redshift bin at $z=0.45$, but the general trends, and the relative advantage of the $6\times2$pt  method with respect to the old CSC method, are very much the same for all redshift bins. As already mentioned, when $r\to0$ one recovers the $3\times2$pt CSC method of \cite{Amendola:2019lvy}. As we see in Figure~\ref{fig:dep-r}, the gain on the relative $H$ error saturates for $r>10$. For such values, the relative error decreases by 13\% with respect to $r\to 0$. Clearly, for very large $r$, the clustering of the standard candles themselves does not add much in terms of strengthening the constraints, and one could have gotten similar results by a $3\times2$pt  method involving only $\delta_{\rm g}$ and $v_{\rm s}$, without the use of $\delta_{\rm s}$. However, this conclusion depends also on the choice of $\beta_{\rm g}$ and $\beta_{\rm s}.$ If we assume $b_{\rm s}$ to be half or double $b_{\rm g}$, then  the inclusion of $\delta_{\rm s}$ can actually improve significantly the constraints, as one can see in Figure~\ref{fig:dep-r}, up to 33\%. This is in fact expected since the multiple tracer method is more effective the more different the tracers' biases are. We discuss this effect more in detail next.

\subsection{Dependence on $b_{\rm g}$ and $b_{\rm s}$}

\begin{figure}
\includegraphics[width=.98\columnwidth]{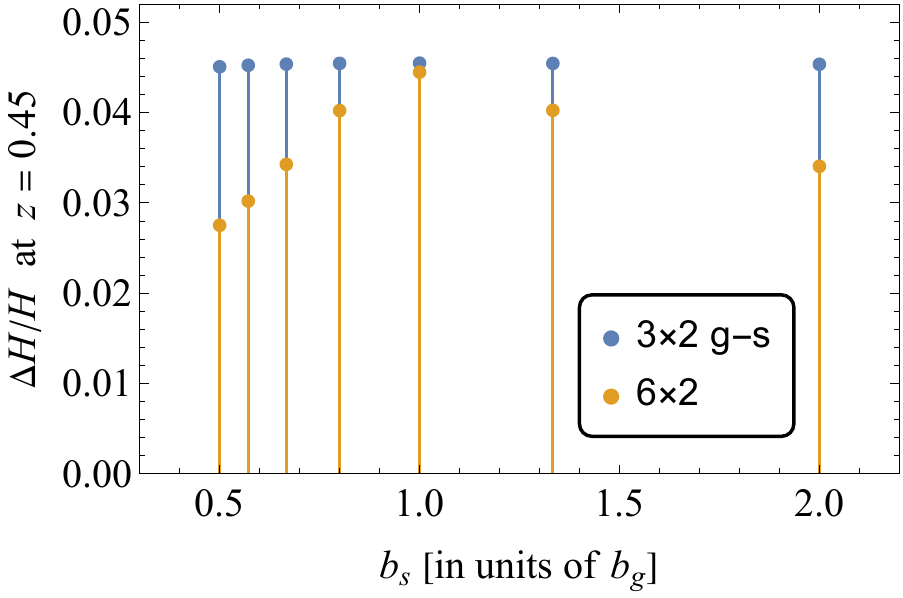}
\caption{Dependence on $b_{\rm g}/b_{\rm s}$ (keeping $b_g$ fixed) of the relative error on $H$ with 50\% prior on   $\sigma_{\rm v,g,s}$ for the $3\times2$ $g$--$s$ case (blue bars) and for the $6\times2$ method (orange bars). We are keeping $b_g$ fixed and vary $b_s$. We fix $n_{\rm g}/n_{\rm s}=10$. Orange bars are slightly displaced for clarity. \label{fig:dep-bias}}
\end{figure}

\begin{figure}
\includegraphics[width=.98\columnwidth]{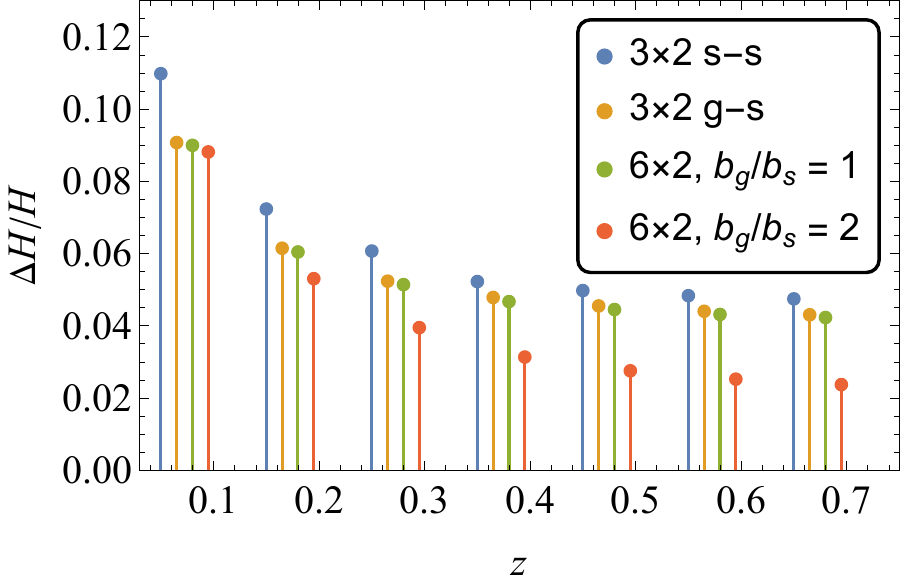}
\caption{Comparing different cases. Bars are slightly displaced from the $z$-bin center for clarity. For the \six\ cases we assumed $n_{\rm g}/n_{\rm s}=10$. \label{fig:dep-cases}}
\end{figure}

In this subsection we fix $b_{\rm g}$ as for the fiducial of Sect.~\ref{sec:surv} and vary $b_{\rm s}$, assuming it to be a fixed fraction of $b_g$ (from now on we put $r=10$). In Figure~\ref{fig:dep-bias} we show the improvement in the $6\times2$pt method when the galaxy and supernovae biases differ (orange bars). As expected, the maximum constraint occur for widely different biases. In particular, for a 50\% prior on $\sigma_{\rm v,g,s}$, we see that the relative error $\Delta H/H$ goes from 4.6\% for equal biases, to 2.8\% when the galaxy bias is twice the supernovae one.  The weakest constraint  occurs, as expected,  for  equal biases (the \six\ appears slightly better simply because in the \three\ case we discard the supernova clustering, which however only contributes by 10\% to the total density; the same occurs in Figure~\ref{fig:dep-cases}). The $6\times2$pt  method clearly improves with respect to the $3\times2$ $g$--$s$ (blue bars) when the biases are different.

\subsection{Comparison among different methods}

In  Figure~\ref{fig:dep-cases} we show the advantages of the \six\ method with respect to various alternatives. The total gain of information for a particular physical quantity when employing a given method~A over a method B can be quantified by a figure-of-merit obtained by multiplying the error gains, i.e.
\begin{equation}
    \prod_i \left(\frac{\rm error\;of\;method\;A}{\rm error\;of\;method\;B}\right)_i\,,
\end{equation}
at every individual redshift shell, where the error here is the uncertainty on $H(z)/H_0$ after marginalizing over all the other quantities.  We find that the $6\times2$pt method  produces   a total gain with respect the $3\times2$~$s$--$s$ of $\,2.7\,$ if $b_g/b_s=1$.  The gain of the $6\times2$pt  method with respect  to the $3\times2$~$g$--$s$, on the other hand, is small because we are taking a large $r=10$ and therefore the clustering of supernovae gives a subdominant contribution, with a marginal total gain of 1.14. If $b_g/b_s$, however, is taken to be 2, then the gain is much larger, around 30 with respect to $3\times2$ $s$--$s$ and 13 with respect to $3\times2$ $g$--$s$.

In Table \ref{tab:mod-ind-fin}, we list the expected relative uncertainty $\Delta H/H$ for the Aggressive and Conservative cases, with the $6\times2$  and the $3\times2$ $g$--$s$ methods\footnote{For the $3\times2$ $g$--$s$  methods, $\beta_s$ is completely degenerate with $P$ and it must therefore be absorbed into its definition.}  (now the galaxy and supernovae biases are as in Sect. \ref{sec:surv}). The total gain of the $6\times2$ over the $3\times2$ $g$--$s$ is 2.8 for the Aggressive case, but just 1.18 for the Conservative one.

Finally, we estimate that the $6\times 2$pt method improves the constraints on $H(z)/H_0$ with respect to the use of the galaxy survey alone (the ``only $g$'' case) by factors of 4 to 1.7  depending on the $z$-bin, with a remarkable  overall gain of roughly 270. This gain depends on the number and size of $z$-bins and on our assumptions that the $z$-bins are independent. If we subdivide the survey into larger bins, the uncertainty decreases but our knowledge of $H(z)$ becomes coarser in $z$-space, and vice versa for smaller bins. At the same time, the relative gain decreases (increases) for larger (smaller) bins, because effectively we have less (more) free parameters. For instance, compressing the data into four bins instead of seven, we find a gain of 15 instead of 270.

Even a slightly higher $k_{\rm max}$ has a large impact on the forecasted constraints. With $k_{\rm max}=0.15$ instead of 0.1 $h$/Mpc, we find that the errors on $H$ improve by 30\% roughly, from 6.7\% at $z=0.05$ down to 2.2\% for the farthest bin.

\begin{table}
\centering
    \setlength{\tabcolsep}{4.9pt}
    \begin{tabular}{|c|c|c|c|c|c|c|}
    \hline
    & \multicolumn{6}{c|}{$\Delta H/H\,\,(\%)$}
    \\
    \cline{2-7}
    \multirow{2}{*}{$z_{\mathrm{bin}}$} & \multicolumn{3}{c|}{Conservative} & \multicolumn{3}{c|}{Aggressive}\\
     & only $g$
     & $3\times2$pt & $6\times2$pt & only $g$ & $3\times2$pt & $6\times2$pt \\
    \hline
    0.05& 83 & 28  & 28 & 39 & 9.1 & 9.0 \\
    0.15& 26 & 14  & 13 & 15 & 6.1 & 6.0 \\
    0.25& 15 & 9.8 & 9.5 & 9.4 & 5.2 & 5.0 \\
    0.35& 14 & 9.1 & 8.2 & 8.5 & 4.7 & 4.0 \\
    0.45&- & -   & -    & 7.0 & 4.5 & 3.6 \\
    0.55&- & -   & -    & 6.1 & 4.4 & 3.3 \\
    0.65& -& -   & -    & 5.5 & 4.3 & 3.2 \\
    \hline
    \end{tabular}
    \caption{\label{tab:mod-ind-fin}  Model-independent marginalized relative $H(z)$ uncertainties for the two survey settings here considered. We take here $n_{\rm g}/n_{\rm s}=10$ and a fiducial $b_{\rm s}=1.0/D_+(z)$. }
\end{table}

\subsection{Constraints on $P(k)$, $\beta(k)$ and $\sigma_{\rm g,s,v}$}

We can also obtain constraints on the other parameters, for instance $P(k,z)$ and $\beta_{\rm s,g}(k,z)$. However it would be very cumbersome and perhaps uninteresting to provide tables for every value of $k$ and $z$, so we select in Table \ref{tab:pbeta} just a few representative cases for the Aggressive survey. The smoothing parameters $\sigma_{\rm g,v,s}$, as expected, are only weakly constrained: their uncertainties  are 28\%, 36\%, 36\%, respectively. Similar values have been obtained with the model dependent approach.

\section{Discussion}

This work shows that the \six\ method enhances the amount of information that can be extracted from the linear regime by combining galaxy clustering with SN clustering and velocities. In particular, it considerably improves upon the standard (\one) approach that employs only galaxy clustering. The \six\ method can be understood as the combination of the multiple tracer and the standard candle clustering techniques. Since typical surveys include many more galaxies than standard candles, the gain with respect to a \three\ method that involves galaxies and standard candles velocities is limited to around $30\%$ in the $\sigma_8, \,\gamma$ FoM. This gain increases the more the galaxy and standard candle biases are different.

Since the peculiar velocity auto power spectrum and  the density-velocity cross power spectrum  have different bias dependencies,  a precise measurement of many bias nuisance parameters becomes possible. Even allowing one free bias parameter in each $\Delta z = 0.1$ redshift bin we could measure them jointly with the cosmological parameters and achieve a relative uncertainty in each bin of only 15\% (5\%) in the Conservative (Aggressive) forecasts for the galaxy biases. This also illustrates the fact that a \six\ analysis in not only more precise but also more robust to uncertainties in the bias modelling.

\begin{table}
    \centering
    \setlength{\tabcolsep}{2.9pt}
    \begin{tabular}{|c|c|c|c|c|c|c|}
    \hline
    \multirow{2}{*}{$z_{\mathrm{bin}}$} & \multicolumn{2}{c|}{$\Delta P/P\,\,(\%)$} & \multicolumn{2}{c|}{$\Delta\beta_{s}/\beta_{s}\,\,(\%)$} & \multicolumn{2}{c|}{$\Delta\beta_{g}/\beta_{g}\,\,(\%)$}\tabularnewline
     & $k=0.1$ & $k=0.05$ & $k=0.1$ & $k=0.05$ & $k=0.1$ & $k=0.05$\tabularnewline
    \hline
    0.25 & 7.7 & 11 & 19 & 23 & 19 & 22\tabularnewline
    0.45 & 5.8 & 8.6 & 16 & 20 & 15 & 19\tabularnewline
    0.65 & 4.7 & 7.4 & 14 & 19 & 14 & 18\tabularnewline
    \hline
    \end{tabular}
    \caption{\label{tab:pbeta}  Relative errors for $P(k,z)$ and $\beta_{\rm g,s}(k,z)$ at some selected values of $k$ and $z$ for the Aggressive survey. $k$-values are in $h$/Mpc units.}
\end{table}

Both the \three\ and \six\ constraints on the growth of structure are very complementary to that from the CMB. The almost perfect orthogonality of the degeneracy directions means that even in our Conservative case the combined  \six\ + CMB results are able to yield improvements on those of the CMB by factors of around 5 in \emph{each} of our cosmological parameters, to wit $\{\sigma_8, \,\gamma, \,h, \,\Omega_{m0},\,\Omega_{k0}\}$. The Aggressive case combined improvements instead are between 8 and 9 for $\sigma_8$, $\gamma$ and $h$.

We also explored and upgraded the model-independent approach first presented in~\cite{Amendola:2019lvy}, in which no assumptions concerning the cosmological model or the bias are needed (while of course still assuming that a linear expansion over a homogeneous and isotropic background is an adequate description of reality). Here too we find important gains. In particular, we have shown that $H(z)/H_0$ can be measured down to 3--4\% for $0.3\le z\le 0.7$, to be compared to 4--5\% for \three\ and 6--8\% when employing only galaxy clustering in the same regime.

We have so far assumed global (i.e. redshift independent) non-linear RSD smoothing factors $\sigma_{\rm g}$, $\sigma_{\rm s}$, and $\sigma_{\rm v}$. If we instead allow them to vary freely in each redshift bin as we already allow for the bias parameters (keeping the same 50\% priors), we find that the Conservative forecasts remain essentially unchanged in both model-dependent and independent approaches. For the Aggressive survey in the model-dependent \six\ (\three), we get 14\% (20\%) larger uncertainties for $\gamma$ and 4\% smaller (1\% larger) for $\sigma_8$ and 2\% (3\%) larger for all $b_g$ parameters. For the \{$\Omega_{m0},\Omega_{k0},\,\,h$\} and for the $b_s$ parameters we get  similar uncertainties. This means that the advantages of \six\ over \three\ are even slightly larger in this case. In the model-independent case \six\ we get on average a 14\% deterioration in precision in $\Delta H/H$, spanning from 1\% at $z=0.05$ to a bit over 20\% for $z\ge0.4$.

In this work we have neglected the effect of redshift errors, which are a crucial aspect of photometric redshift surveys. To compensate for this, we assume a modest number of observed galaxies. We leave an exploration of the regressions involved by adding this source of error, and its possible mitigation by either having a higher number density of galaxies, or by being able to reach higher values of $k_{\rm max}$, for future studies.

The \six\ method might be expanded in a straightforward manner in various directions. First, one should include shear lensing, therefore building a $10\times2$pt correlation matrix. Secondly, one can include other standard candles, in particular gravitational waves. Third, there is the intriguing possibility of measuring the spatial curvature in a model-independent way. These topics will be studied in forthcoming works.

\section*{Acknowledgments}

We thank Adam Mantz and David Rapetti for providing the MCMC chains for their CMB constraints. We also thank Johan Comparat, Will Hartley, Michelle Lochner and Emiliano Sefusatti for useful discussions and suggestions. We acknowledge interesting suggestions from the anonymous referee which improved the presentation of this manuscript. MQ is supported by the Brazilian research agencies CNPq, FAPERJ and CAPES. LA acknowledges support from DFG project  456622116. BM acknowledges support from the Brazilian research agency FAPERJ.
This study was financed in part by the Coordenação de Aperfeiçoamento de Pessoal de Nível Superior - Brasil (CAPES) - Finance Code 001. We acknowledge support from the CAPES-DAAD bilateral project  ``Data Analysis and Model Testing in the Era of Precision Cosmology''.

\section*{Data availability}

The data underlying this article will be shared on reasonable request to the corresponding author. The Mathematica notebooks which can be used to reproduce the FM  calculations can be downloaded at \url{www.github.com/mquartin/clustering} (for the model dependent case) and \url{www.github.com/itpamendola/clustering} (for the model-independent case).

\bibliography{references,scaling_bib,refs_bruno}

\begin{thebibliography}{}

\bibitem[\protect\citeauthoryear{Abbott et~al.,}{Abbott
  et~al.}{2022}]{DES:2021wwk}
Abbott T. M.~C.,  et~al., 2022, Phys. Rev. D, 105, 023520, \eprint{2105.13549}

\bibitem[\protect\citeauthoryear{Abramo}{Abramo}{2011}]{Abramo:2012}
Abramo L.~R.,  2011, MNRAS 420, 3, pp. 2032 (2012), \eprint{1108.5449}

\bibitem[\protect\citeauthoryear{Abramo \& Amendola}{Abramo \&
  Amendola}{2019}]{Abramo:2019ejj}
Abramo L.~R.,  Amendola L.,  2019, JCAP, 1906, 030, \eprint{1904.00673}

\bibitem[\protect\citeauthoryear{Abramo \& Leonard}{Abramo \&
  Leonard}{2013}]{Abramo:2013}
Abramo L.~R.,  Leonard K.~E.,  2013, MNRAS, 432, 318, \eprint{1302.5444}

\bibitem[\protect\citeauthoryear{Alcock \& Paczynski}{Alcock \&
  Paczynski}{1979}]{Alcock:1979mp}
Alcock C.,  Paczynski B.,  1979, Nature, 281, 358

\bibitem[\protect\citeauthoryear{Amara \& Refregier}{Amara \&
  Refregier}{2008}]{Amara:2007as}
Amara A.,  Refregier A.,  2008, MNRAS, 391, 228, \eprint{0710.5171}

\bibitem[\protect\citeauthoryear{Amendola \& Quartin}{Amendola \&
  Quartin}{2021}]{Amendola:2019lvy}
Amendola L.,  Quartin M.,  2021, MNRAS, 504, 3884, \eprint{1912.10255}

\bibitem[\protect\citeauthoryear{Amendola \& Quercellini}{Amendola \&
  Quercellini}{2004}]{Amendola_2004}
Amendola L.,  Quercellini C.,  2004, Phys. Rev. Lett., 92, 181102,
  \eprint{astro-ph/0403019}

\bibitem[\protect\citeauthoryear{Amendola, Quercellini \& Giallongo}{Amendola
  et~al.}{2005}]{Amendola:2004be}
Amendola L.,  Quercellini C.,    Giallongo E.,  2005, MNRAS, 357, 429,
  \eprint{astro-ph/0404599}

\bibitem[\protect\citeauthoryear{Asgari et~al.,}{Asgari
  et~al.}{2021}]{KiDS:2020suj}
Asgari M.,  et~al., 2021, A\&A, 645, A104, \eprint{2007.15633}

\bibitem[\protect\citeauthoryear{{Avelino}, {Friedman}, {Mandel}, {Jones},
  {Challis} \& {Kirshner}}{{Avelino} et~al.}{2019}]{Avelino2019}
{Avelino} A.,  {Friedman} A.~S.,  {Mandel} K.~S.,  {Jones} D.~O.,  {Challis}
  P.~J.,    {Kirshner} R.~P.,  2019, ApJ, 887, 106, \eprint{1902.03261},
  \adsurl{https://ui.adsabs.harvard.edu/abs/2019ApJ...887..106A}

\bibitem[\protect\citeauthoryear{{Ballinger}, {Peacock} \&
  {Heavens}}{{Ballinger} et~al.}{1996}]{1996MNRAS.282..877B}
{Ballinger} W.~E.,  {Peacock} J.~A.,    {Heavens} A.~F.,  1996, MNRAS, 282,
  877, \eprint{astro-ph/9605017},
  \adsurl{https://ui.adsabs.harvard.edu/abs/1996MNRAS.282..877B}

\bibitem[\protect\citeauthoryear{Betoule et~al.,}{Betoule
  et~al.}{2014}]{Betoule:2014frx}
Betoule M.,  et~al., 2014, Astron.Astrophys., 568, A22, \eprint{1401.4064}

\bibitem[\protect\citeauthoryear{Bonoli et~al.,}{Bonoli
  et~al.}{2021}]{Bonoli:2020ciz}
Bonoli S.,  et~al., 2021, A\&A, 653, A31, \eprint{2007.01910}

\bibitem[\protect\citeauthoryear{Boruah, Hudson \& Lavaux}{Boruah
  et~al.}{2020}]{Boruah:2019icj}
Boruah S.~S.,  Hudson M.~J.,    Lavaux G.,  2020, MNRAS, 498, 2703,
  \eprint{1912.09383}

\bibitem[\protect\citeauthoryear{Burkey \& Taylor}{Burkey \&
  Taylor}{2004}]{Burkey:2003rk}
Burkey D.,  Taylor A.~N.,  2004, MNRAS, 347, 255, \eprint{astro-ph/0310912}

\bibitem[\protect\citeauthoryear{Cappellaro et~al.,}{Cappellaro
  et~al.}{2015}]{Cappellaro:2015}
Cappellaro E.,  et~al., 2015, A\&A, 584, A62, \eprint{1509.04496}

\bibitem[\protect\citeauthoryear{{Carlberg}, {Sullivan}, {Le Borgne}, {Conley},
  {Howell} et~al.,}{{Carlberg} et~al.}{2008}]{Carlberg2008}
{Carlberg} R.~G.,  {Sullivan} M.,  {Le Borgne} D.,  {Conley} A.,  {Howell}
  D.~A.,    et~al., 2008, ApJ, 682, L25, \eprint{0805.3983},
  \adsurl{https://ui.adsabs.harvard.edu/abs/2008ApJ...682L..25C}

\bibitem[\protect\citeauthoryear{{Castro} \& {Quartin}}{{Castro} \&
  {Quartin}}{2014}]{Castro:2014oja}
{Castro} T.,  {Quartin} M.,  2014, MNRAS, 443, L6, \eprint{1403.0293},
  \adsurl{http://adsabs.harvard.edu/abs/2014MNRAS.443L...6C}

\bibitem[\protect\citeauthoryear{Castro, Quartin \& Benitez-Herrera}{Castro
  et~al.}{2016}]{Castro:2015rrx}
Castro T.,  Quartin M.,    Benitez-Herrera S.,  2016, Phys. Dark Univ., 13, 66,
  \eprint{1511.08695}

\bibitem[\protect\citeauthoryear{Dam, Bolejko \& Lewis}{Dam
  et~al.}{2021}]{Dam:2021fff}
Dam L.,  Bolejko K.,    Lewis G.~F.,  2021, JCAP, 09, 018, \eprint{2105.12933}

\bibitem[\protect\citeauthoryear{{Das}, {Louis}, {Nolta}, {Addison},
  {Battistelli} et~al.,}{{Das} et~al.}{2014}]{Das:2013zf}
{Das} S.,  {Louis} T.,  {Nolta} M.~R.,  {Addison} G.~E.,  {Battistelli} E.~S.,
    et~al., 2014, JCAP, 04, 014, \eprint{1301.1037}

\bibitem[\protect\citeauthoryear{Davis et~al.,}{Davis
  et~al.}{2011}]{Davis:2010jq}
Davis T.~M.,  et~al., 2011, ApJ, 741, 67, \eprint{1012.2912}

\bibitem[\protect\citeauthoryear{{DESI Collaboration}}{{DESI
  Collaboration}}{2016}]{desicollaboration2016}
{DESI Collaboration}, 2016, The DESI Experiment Part I: Science,Targeting, and
  Survey Design, \eprint{1611.00036}

\bibitem[\protect\citeauthoryear{{eBOSS Collaboration}}{{eBOSS
  Collaboration}}{2021}]{eBOSS:2020yzd}
{eBOSS Collaboration} 2021, Phys. Rev. D, 103, 083533, \eprint{2007.08991}

\bibitem[\protect\citeauthoryear{{Euclid Collaboration}}{{Euclid
  Collaboration}}{2018}]{Amendola:2016saw}
{Euclid Collaboration} 2018, Living Rev. Rel., 21, 2, \eprint{1606.00180}

\bibitem[\protect\citeauthoryear{Foroozan, Krolewski \& Percival}{Foroozan
  et~al.}{2021}]{Foroozan:2021zzu}
Foroozan S.,  Krolewski A.,    Percival W.~J.,  2021, JCAP, 10, 044,
  \eprint{2106.11432}

\bibitem[\protect\citeauthoryear{Garcia, Quartin \& Siffert}{Garcia
  et~al.}{2020}]{Garcia:2019ita}
Garcia K.,  Quartin M.,    Siffert B.~B.,  2020, Phys. Dark Univ., 29, 100519,
  \eprint{1905.00746}

\bibitem[\protect\citeauthoryear{Gordon, Land \& Slosar}{Gordon
  et~al.}{2007}]{Gordon:2007zw}
Gordon C.,  Land K.,    Slosar A.,  2007, Phys. Rev. Lett., 99, 081301,
  \eprint{0705.1718}

\bibitem[\protect\citeauthoryear{Graziani et~al.,}{Graziani
  et~al.}{2020}]{Graziani:2020kkr}
Graziani R.,  et~al., 2020, \eprint{2001.09095}

\bibitem[\protect\citeauthoryear{Grillo et~al.,}{Grillo
  et~al.}{2018}]{Grillo:2018ume}
Grillo C.,  et~al., 2018, ApJ, 860, 94, \eprint{1802.01584}

\bibitem[\protect\citeauthoryear{Grillo, Rosati, Suyu, Caminha, Mercurio \&
  Halkola}{Grillo et~al.}{2020}]{Grillo:2020yvj}
Grillo C.,  Rosati P.,  Suyu S.~H.,  Caminha G.~B.,  Mercurio A.,    Halkola
  A.,  2020, ApJ, 898, 87, \eprint{2001.02232}

\bibitem[\protect\citeauthoryear{{Guy}, {Sullivan}, {Conley}, {Regnault},
  {Astier} et~al.,}{{Guy} et~al.}{2010}]{Guy2010}
{Guy} J.,  {Sullivan} M.,  {Conley} A.,  {Regnault} N.,  {Astier} P.,
  et~al., 2010, A\&A, 523, A7, \eprint{1010.4743},
  \adsurl{https://ui.adsabs.harvard.edu/abs/2010A&A...523A...7G}

\bibitem[\protect\citeauthoryear{Howlett, Robotham, Lagos \& Kim}{Howlett
  et~al.}{2017}]{Howlett:2017asw}
Howlett C.,  Robotham A. S.~G.,  Lagos C. D.~P.,    Kim A.~G.,  2017, ApJ, 847,
  128, \eprint{1708.08236}

\bibitem[\protect\citeauthoryear{Howlett, Staveley-Smith \& Blake}{Howlett
  et~al.}{2017}]{Howlett:2016urc}
Howlett C.,  Staveley-Smith L.,    Blake C.,  2017, MNRAS, 464, 2517,
  \eprint{1609.08247}

\bibitem[\protect\citeauthoryear{{Hui} \& {Greene}}{{Hui} \&
  {Greene}}{2006}]{hui2006}
{Hui} L.,  {Greene} P.~B.,  2006, \prd, 73, 123526, \eprint{astro-ph/0512159},
  \adsurl{http://adsabs.harvard.edu/abs/2006PhRvD..73l3526H}

\bibitem[\protect\citeauthoryear{Huterer, Shafer, Scolnic \& Schmidt}{Huterer
  et~al.}{2017}]{Huterer:2016uyq}
Huterer D.,  Shafer D.,  Scolnic D.,    Schmidt F.,  2017, JCAP, 1705, 015,
  \eprint{1611.09862}

\bibitem[\protect\citeauthoryear{Huterer \& Takada}{Huterer \&
  Takada}{2005}]{Huterer:2004tr}
Huterer D.,  Takada M.,  2005, Astropart. Phys., 23, 369,
  \eprint{astro-ph/0412142}

\bibitem[\protect\citeauthoryear{Johnson et~al.,}{Johnson
  et~al.}{2014}]{Johnson:2014kaa}
Johnson A.,  et~al., 2014, MNRAS, 444, 3926, \eprint{1404.3799}

\bibitem[\protect\citeauthoryear{{J{\"o}nsson}, Sullivan, Hook, Basa, Carlberg
  et~al.,}{{J{\"o}nsson} et~al.}{2010}]{Jonsson:2010wx}
{J{\"o}nsson} J.,  Sullivan M.,  Hook I.,  Basa S.,  Carlberg R.,    et~al.,
  2010, Mon.Not.Roy.Astron.Soc., 405, 535, \eprint{1002.1374}

\bibitem[\protect\citeauthoryear{{Kaiser}}{{Kaiser}}{1987}]{Kaiser1987}
{Kaiser} N.,  1987, MNRAS, 227, 1,
  \adsurl{http://adsabs.harvard.edu/abs/1987MNRAS.227....1K}

\bibitem[\protect\citeauthoryear{Kase \& Tsujikawa}{Kase \&
  Tsujikawa}{2020}]{RyotaroTsujikawa2020a}
Kase R.,  Tsujikawa S.,  2020, JCAP, 11, 032, \eprint{2005.13809}

\bibitem[\protect\citeauthoryear{{Koda}, {Blake}, {Davis}, {Magoulas},
  {Springob}, {Scrimgeour}, {Johnson}, {Poole} \& {Staveley-Smith}}{{Koda}
  et~al.}{2014}]{2014MNRAS.445.4267K}
{Koda} J.,  {Blake} C.,  {Davis} T.,  {Magoulas} C.,  {Springob} C.~M.,
  {Scrimgeour} M.,  {Johnson} A.,  {Poole} G.~B.,    {Staveley-Smith} L.,
  2014, MNRAS, 445, 4267, \eprint{1312.1022},
  \adsurl{https://ui.adsabs.harvard.edu/abs/2014MNRAS.445.4267K}

\bibitem[\protect\citeauthoryear{Laureijs et~al.,}{Laureijs
  et~al.}{2011}]{Laureijs:2011:1}
Laureijs R.,  et~al., 2011, \eprint{1110.3193}

\bibitem[\protect\citeauthoryear{Lewis, Challinor \& Lasenby}{Lewis
  et~al.}{2000}]{Lewis:1999bs}
Lewis A.,  Challinor A.,    Lasenby A.,  2000, ApJ, 538, 473,
  \eprint{astro-ph/9911177}

\bibitem[\protect\citeauthoryear{Libanore, Artale, Karagiannis, Liguori,
  Bartolo, Bouffanais, Mapelli \& Matarrese}{Libanore
  et~al.}{2022}]{Libanore:2021jqv}
Libanore S.,  Artale M.~C.,  Karagiannis D.,  Liguori M.,  Bartolo N.,
  Bouffanais Y.,  Mapelli M.,    Matarrese S.,  2022, JCAP, 02, 003,
  \eprint{2109.10857}

\bibitem[\protect\citeauthoryear{{Linder}}{{Linder}}{2005}]{Linder2005}
{Linder} E.~V.,  2005, \prd, 72, 043529, \eprint{astro-ph/0507263},
  \adsurl{https://ui.adsabs.harvard.edu/abs/2005PhRvD..72d3529L}

\bibitem[\protect\citeauthoryear{{Linder} \& {Cahn}}{{Linder} \&
  {Cahn}}{2007}]{LinderCahn2007}
{Linder} E.~V.,  {Cahn} R.~N.,  2007, Astroparticle Physics, 28, 481,
  \eprint{astro-ph/0701317},
  \adsurl{https://ui.adsabs.harvard.edu/abs/2007APh....28..481L}

\bibitem[\protect\citeauthoryear{Lochner, McEwen, Peiris, Lahav \&
  Winter}{Lochner et~al.}{2016}]{Lochner:2016hbn}
Lochner M.,  McEwen J.~D.,  Peiris H.~V.,  Lahav O.,    Winter M.~K.,  2016,
  ApJS, 225, 31, \eprint{1603.00882}

\bibitem[\protect\citeauthoryear{{LSST Science Collaborations}}{{LSST Science
  Collaborations}}{2009}]{LSSTScience:2009jmu}
{LSST Science Collaborations} 2009, \eprint{0912.0201}

\bibitem[\protect\citeauthoryear{Macaulay, Davis, Scovacricchi, Bacon, Collett
  \& Nichol}{Macaulay et~al.}{2017}]{Macaulay:2016uwy}
Macaulay E.,  Davis T.~M.,  Scovacricchi D.,  Bacon D.,  Collett T.~E.,
  Nichol R.~C.,  2017, MNRAS, 467, 259, \eprint{1607.03966}

\bibitem[\protect\citeauthoryear{Macaulay et~al.,}{Macaulay
  et~al.}{2020}]{DES:2020kbf}
Macaulay E.,  et~al., 2020, MNRAS, 496, 4051, \eprint{2007.07956}

\bibitem[\protect\citeauthoryear{McDonald \& Seljak}{McDonald \&
  Seljak}{2009}]{McDonald:2009}
McDonald P.,  Seljak U.,  2009, JCAP, 0910, 007, \eprint{0810.0323}

\bibitem[\protect\citeauthoryear{{Magira}, {Jing} \& {Suto}}{{Magira}
  et~al.}{2000}]{2000ApJ...528...30M}
{Magira} H.,  {Jing} Y.~P.,    {Suto} Y.,  2000, ApJ, 528, 30,
  \eprint{astro-ph/9907438},
  \adsurl{https://ui.adsabs.harvard.edu/abs/2000ApJ...528...30M}

\bibitem[\protect\citeauthoryear{Mantz et~al.,}{Mantz
  et~al.}{2015}]{Mantz:2014paa}
Mantz A.~B.,  et~al., 2015, MNRAS, 446, 2205, \eprint{1407.4516}

\bibitem[\protect\citeauthoryear{March, Trotta, Amendola \& Huterer}{March
  et~al.}{2011}]{March:2011rv}
March M.,  Trotta R.,  Amendola L.,    Huterer D.,  2011,
  Mon.Not.Roy.Astron.Soc., 415, 143, \eprint{1101.1521}

\bibitem[\protect\citeauthoryear{{Mukherjee} \& {Wandelt}}{{Mukherjee} \&
  {Wandelt}}{2018}]{MukherjeeWandelt2018}
{Mukherjee} S.,  {Wandelt} B.~D.,  2018, arXiv e-prints, p. arXiv:1808.06615,
  \eprint{1808.06615},
  \adsurl{https://ui.adsabs.harvard.edu/abs/2018arXiv180806615M}

\bibitem[\protect\citeauthoryear{Palmese \& Kim}{Palmese \&
  Kim}{2021}]{Palmese:2020kxn}
Palmese A.,  Kim A.~G.,  2021, Phys. Rev. D, 103, 103507, \eprint{2005.04325}

\bibitem[\protect\citeauthoryear{{Peebles}}{{Peebles}}{1980}]{Peebles1980}
{Peebles} P.~J.~E.,  1980, {The large-scale structure of the universe},
  \adsurl{https://ui.adsabs.harvard.edu/abs/1980lssu.book.....P.
}

\bibitem[\protect\citeauthoryear{{Percival} \& {White}}{{Percival} \&
  {White}}{2009}]{PercivalWhite2009}
{Percival} W.~J.,  {White} M.,  2009, MNRAS, 393, 297, \eprint{0808.0003},
  \adsurl{https://ui.adsabs.harvard.edu/abs/2009MNRAS.393..297P}

\bibitem[\protect\citeauthoryear{Peterson et~al.,}{Peterson
  et~al.}{2021}]{Peterson:2021hel}
Peterson E.~R.,  et~al., 2021, \eprint{2110.03487}

\bibitem[\protect\citeauthoryear{{Planck Collaboration VI}}{{Planck
  Collaboration VI}}{2020}]{Planck:2018vyg}
{Planck Collaboration VI} 2020, A\&A, 641, A6, \eprint{1807.06209}

\bibitem[\protect\citeauthoryear{Quartin, Marra \& Amendola}{Quartin
  et~al.}{2014}]{Quartin:2013moa}
Quartin M.,  Marra V.,    Amendola L.,  2014, Phys.Rev., D89, 023009,
  \eprint{1307.1155}

\bibitem[\protect\citeauthoryear{{Rapetti}, {Allen}, {Mantz} \&
  {Ebeling}}{{Rapetti} et~al.}{2009}]{Rapetti2009}
{Rapetti} D.,  {Allen} S.~W.,  {Mantz} A.,    {Ebeling} H.,  2009, MNRAS, 400,
  699, \eprint{0812.2259},
  \adsurl{https://ui.adsabs.harvard.edu/abs/2009MNRAS.400..699R}

\bibitem[\protect\citeauthoryear{{Richard}, {Kneib}, {Blake}, {Raichoor},
  {Comparat} et~al.,}{{Richard} et~al.}{2019}]{Richard2019}
{Richard} J.,  {Kneib} J.~P.,  {Blake} C.,  {Raichoor} A.,  {Comparat} J.,
  et~al., 2019, The Messenger, 175, 50, \eprint{1903.02474},
  \adsurl{https://ui.adsabs.harvard.edu/abs/2019Msngr.175...50R}

\bibitem[\protect\citeauthoryear{Riess, Casertano, Yuan, Bowers, Macri, Zinn \&
  Scolnic}{Riess et~al.}{2021}]{Riess:2020fzl}
Riess A.~G.,  Casertano S.,  Yuan W.,  Bowers J.~B.,  Macri L.,  Zinn J.~C.,
  Scolnic D.,  2021, Astrophys. J. Lett., 908, L6, \eprint{2012.08534}

\bibitem[\protect\citeauthoryear{Samushia et~al.,}{Samushia
  et~al.}{2011}]{Samushia:2010ki}
Samushia L.,  et~al., 2011, MNRAS, 410, 1993, \eprint{1006.0609}

\bibitem[\protect\citeauthoryear{Sanchez}{Sanchez}{2020}]{Sanchez:2020vvb}
Sanchez A.~G.,  2020, Phys. Rev. D, 102, 123511, \eprint{2002.07829}

\bibitem[\protect\citeauthoryear{Scovacricchi, Nichol, Macaulay \&
  Bacon}{Scovacricchi et~al.}{2017}]{Scovacricchi:2016ylt}
Scovacricchi D.,  Nichol R.~C.,  Macaulay E.,    Bacon D.,  2017, MNRAS, 465,
  2862, \eprint{1611.01315}

\bibitem[\protect\citeauthoryear{Seljak}{Seljak}{2009}]{Seljak:2009}
Seljak U.,  2009, Phys. Rev. Lett., 102, 021302, \eprint{0807.1770}

\bibitem[\protect\citeauthoryear{Seo \& Eisenstein}{Seo \&
  Eisenstein}{2003}]{Seo:2003pu}
Seo H.-J.,  Eisenstein D.~J.,  2003, ApJ, 598, 720, \eprint{astro-ph/0307460}

\bibitem[\protect\citeauthoryear{{Song} \& {Percival}}{{Song} \&
  {Percival}}{2009}]{SongPercival2009}
{Song} Y.-S.,  {Percival} W.~J.,  2009, \jcap, 2009, 004, \eprint{0807.0810},
  \adsurl{https://ui.adsabs.harvard.edu/abs/2009JCAP...10..004S}

\bibitem[\protect\citeauthoryear{Story et~al.,}{Story
  et~al.}{2013}]{Story:2012wx}
Story K.~T.,  et~al., 2013, ApJ, 779, 86, \eprint{1210.7231}

\bibitem[\protect\citeauthoryear{Takahashi, Sato, Nishimichi, Taruya \&
  Oguri}{Takahashi et~al.}{2012}]{Takahashi:2012em}
Takahashi R.,  Sato M.,  Nishimichi T.,  Taruya A.,    Oguri M.,  2012, ApJ,
  761, 152, \eprint{1208.2701}

\bibitem[\protect\citeauthoryear{Tegmark}{Tegmark}{1997}]{Tegmark:1997rp}
Tegmark M.,  1997, Phys. Rev. Lett., 79, 3806, \eprint{astro-ph/9706198}

\bibitem[\protect\citeauthoryear{Tr\"oster et~al.,}{Tr\"oster
  et~al.}{2021}]{KiDS:2020ghu}
Tr\"oster T.,  et~al., 2021, A\&A, 649, A88, \eprint{2010.16416}

\bibitem[\protect\citeauthoryear{Tsaprazi et~al.,}{Tsaprazi
  et~al.}{2021}]{Tsaprazi2021}
Tsaprazi E.,  et~al., 2021, MNRAS, 510, 366, \eprint{2109.02651}

\bibitem[\protect\citeauthoryear{Vargas~dos Santos, Quartin \& Reis}{Vargas~dos
  Santos et~al.}{2019}]{VargasdosSantos:2019ovq}
Vargas~dos Santos M.,  Quartin M.,    Reis R. R.~R.,  2019, MNRAS,
  \eprint{1908.04210}

\bibitem[\protect\citeauthoryear{{White}, {Song} \& {Percival}}{{White}
  et~al.}{2009}]{WhiteSongPercival2009}
{White} M.,  {Song} Y.-S.,    {Percival} W.~J.,  2009, MNRAS, 397, 1348,
  \eprint{0810.1518},
  \adsurl{https://ui.adsabs.harvard.edu/abs/2009MNRAS.397.1348W}

\bibitem[\protect\citeauthoryear{Zheng, Zhang \& Jing}{Zheng
  et~al.}{2015}]{Zheng:2014vla}
Zheng Y.,  Zhang P.,    Jing Y.,  2015, Phys. Rev., D91, 123512,
  \eprint{1410.1256}

\bibitem[\protect\citeauthoryear{{Zhou}, {Newman}, {Mao}, {Meisner},
  {Moustakas} et~al.,}{{Zhou} et~al.}{2021}]{Zhou2021}
{Zhou} R.,  {Newman} J.~A.,  {Mao} Y.-Y.,  {Meisner} A.,  {Moustakas} J.,
  et~al., 2021, MNRAS, 501, 3309, \eprint{2001.06018},
  \adsurl{https://ui.adsabs.harvard.edu/abs/2021MNRAS.501.3309Z}

\end{thebibliography}




\appendix

\section{List of derivatives}\label{app:deriv}

Here we show all the different relevant derivatives that appear in the model-dependent case due to the Alcock-Paczynski terms. All derivatives are naturally taken around the reference cosmology, which simplify the equations. These equations are similar to the ones found in~\cite{Samushia:2010ki}. As usual, we define $E(z) \equiv H(z) / H_0$ and $H_0 = 100 h$ km/s/Mpc.
\begin{align}
    \frac{\partial \ln \Upsilon}{\partial \Omega_{m0}} & = F_3(z) - 2 C_3(z) \,, \\
    \frac{\partial P_{\rm mm}(k)}{\partial \Omega_{m0}} & = \frac{\partial P_{\rm mm}(k_r)}{\partial \Omega_{m0}}  + \frac{\partial P_{\rm mm}}{\partial \ln k} \frac{\partial \ln k}{\partial \Omega_{m0}}  \,, \\
    \frac{\partial \ln k}{\partial \Omega_{m0}} & = \mu^2 F_3(z) + (\mu^2-1) C_3(z) \,, \\
    \frac{\partial \mu}{\partial \Omega_{m0}} & = \mu(1-\mu^2)\big(F_3(z) + C_3(z)\big) \,, \\
    \frac{\partial \ln \Upsilon}{\partial \Omega_{k0}} & = F_2(z) -2 C_2(z) \,, \\
    \frac{\partial P_{\rm mm}(k)}{\partial \Omega_{k0}} & = \frac{\partial P_{\rm mm}(k_r)}{\partial \Omega_{k0}}  + \frac{\partial P_{\rm mm}}{\partial \ln k} \frac{\partial \ln k}{\partial \Omega_{k0}}  \,, \\
    \frac{\partial \ln k}{\partial \Omega_{k0}} & = \mu^2 F_2(z) + (\mu^2-1) C_2(z) \,, \\
    \frac{\partial \mu}{\partial \Omega_{k0}} & = \mu(1-\mu^2)\big(F_2(z) + C_2(z)\big) \,, \\
    \frac{\partial \ln \Upsilon}{\partial h} & = \frac{3}{h} \,, \\
    \frac{\partial P_{\rm mm}(k)}{\partial h} & = \frac{\partial P_{\rm mm}(k_r)}{\partial h}  + \frac{\partial P_{\rm mm}}{\partial \ln k} \frac{\partial \ln k}{\partial h}  \,. \\
    \frac{\partial \mu}{\partial h} & = \frac{\partial \eta}{\partial h} = 0 \,, \\
    \frac{\partial \ln k}{\partial h} & = \frac{1}{h}\,, \\
    F_n(z) &\equiv \frac{(1+z)^n -1}{2 E^2(z)} ,\\
    C_3(z) &\equiv -\frac{\int \dd z' \frac{ F_3(z')}{E(z')}}{\int \dd z' \frac{1}{E(z')}} \,, \\
    C_2(z) &\equiv -\frac{\int \dd z' \frac{ F_2(z')}{E(z')}}{\int \dd z' \frac{1}{E(z')}} + \frac{1}{6} \left[\int \dd z' \frac{1}{E(z')}\right]^2 \,.
\end{align}

The partial derivatives with respect to the background cosmology parameters ($\Omega_{m0}$, $\Omega_{k0}$ and $h$) are all taken at fixed $\sigma_8$ values. This is done in order to keep all parameters independent in our FM analysis, but requires varying the amplitude $A_s$ when any one of these parameters change in order to keep $\sigma_8$ fixed. Note that this means that in this case the overall effect of $h$ in $P_{mm}$ becomes less intuitive~\citep{Sanchez:2020vvb}.

The derivatives which depend on the shape of the power spectrum were computed in practice by running CAMB at slightly higher and lower parameter values (compared to the fiducial ones) and computing finite differences.

\section{The cross-spectrum noise term}\label{app:cross}

The Fourier transform of a discrete set with density $n_0=N/V$ is
\begin{align}
    \delta_{k} & =\frac{1}{V}\int\delta(x)e^{ikx}d^{3}x\\
     & =\frac{1}{V}\int\sum_{i}\frac{\delta_{D}(x-x_{i})}{n_{0}}e^{ikx}d^{3}x\\
     & =\frac{1}{V}\sum_{i}\frac{1}{n_{0}}e^{ikx_{i}}=\sum_{i}\frac{1}{N}e^{ikx_{i}}\,,
\end{align}
where $\delta_D$ is Dirac's delta. The power spectrum is usually defined with an overall volume factor, so
\begin{align}
    P & \,=\,V\delta_{k}\delta_{k}^{*}\\
     & \,=\,\frac{V}{N^{2}}\sum_{i,j}e^{ik(x_{i}-x_{j})}\,.
\end{align}
If we have two populations (subscripts $1,2$) in the same volume, with partial or total overlap,
then
\begin{align}
    P_{12} & =\sum_{i}\frac{1}{N_{1}}e^{ikx_{i}}\sum_{j}\frac{1}{N_{2}}e^{ikx_{j}} =\frac{V}{N_{1}N_{2}}\sum_{i,j}e^{ik(x_{i}-x_{j})}\\
    & =\frac{V}{N_{1}N_{2}}\sum_{i\not=j}e^{ik(x_{i}-x_{j})}+\frac{V}{N_{1}N_{2}}\sum_{i=j}1\\
    & =\frac{V}{N_{1}N_{2}}\sum_{i\not=j}e^{ik(x_{i}-x_{j})}+\frac{V}{N_{1}N_{2}}N_{12} \,,
\end{align}
where $N_{12}$ is the number of points that are both of type 1 and
type 2. The last term is the mixed shot-noise that appears in Eq. (\ref{eq:pgs}):
\begin{equation}
    P_{\rm shot}=\frac{V}{N_{1}N_{2}}N_{12}=\frac{V^{2}}{N_{1}N_{2}}\frac{N_{12}}{V}
    =\frac{n_{12}}{n_{1}n_{2}} \,.
\end{equation}

\section{Combining \six\ and CMB constrains}\label{app:cmb-combination}

\cite{Mantz:2014paa} used Planck 2013 temperature data together with WMAP polarization and ACT and SPT data to find constraints on $\sigma_8$ and $\gamma$ assuming a flat $\Lambda$CDM background. These CMB constraints on $\gamma$ come mostly from the ISW effect. In order to combine these constraints with those from the \six\ method, we multiplied both posteriors in ${\sigma_8, \gamma}$ after marginalizing over the other parameters. In principle marginalization should instead be performed after the multiplication, but the available CMB results do not include curvature, and the correlation between $\{\Omega_{m0}, \,h\}$ and ${\sigma_8, \gamma}$ is small. In fact, we tested explicitly that multiplying first and subsequently marginalizing over either $\Omega_{m0}$ or $h$ did not change the contours in ${\sigma_8, \gamma}$.

\begin{figure}
    \centering
    \includegraphics[width=.93\columnwidth]{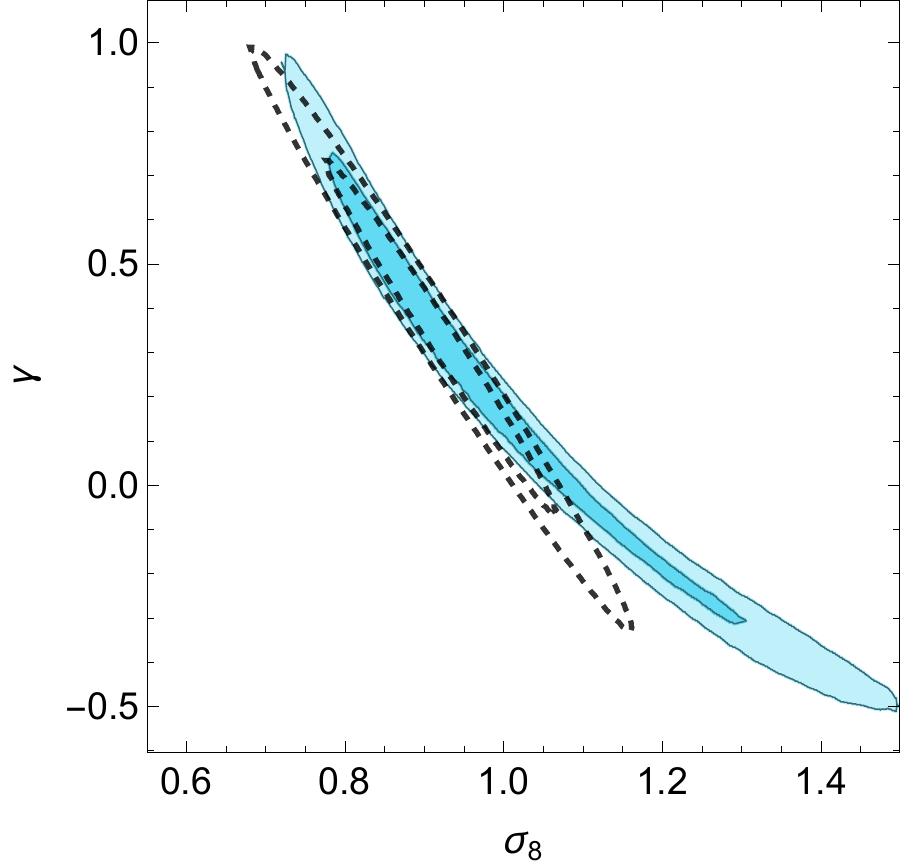}
    \caption{CMB real data constraints on $\{\sigma_8,\,\gamma\}$ and simple Gaussian approximation in the region of interest. \label{fig:cmb-approx}
    }
\end{figure}

Since the \cite{Mantz:2014paa} CMB results for $\gamma$ do not include curvature, they greatly underestimate the uncertainties in $\Omega_{m0}$ and $h$ for the curved case (by over a factor of 3 in each). Thus in order to better compare with the \six\ forecasts in Table~\ref{tab:1Derrors} (and only for this Table) we combined the above Planck 2013 + WMAP polarization covariance on $\{\sigma_8, \,\gamma\}$ with  Planck 2018 TTTEEE covariance matrix (without lensing) for the curved $\Lambda$CDM case on the variables $\{h, \,\Omega_{m0},\,\Omega_{k0}\}$ assuming as an approximation zero correlation between these variables. Assuming flatness the CMB chains confirm this is a good approximation. Planck 2018 TTTEEE constraints on the full $\{\sigma_8, \,\gamma, \,h, \,\Omega_{m0},\,\Omega_{k0}\}$ parameter set would allow to test if the correlations with $\Omega_{k0}$ are also approximately zero, but we are not aware of such an analysis in the current literature.

Adding CMB lensing improves the precision in $\Omega_{m0}$ and $h$ by a factor of around 2.5 and 1.5, respectively, but since we are also not taking into account lensing in our \six\ forecasts, we consider the CMB results without lensing a better benchmark for comparison.

Finally, we also derived a Gaussian approximation to these real data CMB constraints which are a very good fit for $\sigma_8 < 1.1$ and $\gamma>0$. This allows to compute a CMB FM and use all the FM tools directly also when combining results with the CMB, which might be useful to quickly compute further joint CMB constraints. The CMB FM in $\{\sigma_8, \gamma\}$ is given by
\begin{equation}
    F_{\rm CMB} = \begin{pmatrix}
    5341 & 1921 \\
    1921 & 705
    \end{pmatrix}\,,
\end{equation}
and the bestfit is $\{\sigma_8, \gamma\} = \{0.92, \,0.33\}$. Figure~\ref{fig:cmb-approx} illustrates the full CMB constraints and our Gaussian approximation. The range of values in which the CMB posterior is well approximated by its FM contours is the one relevant when combining it with the \six\ results. Using either this Gaussian approximation or the full chains when combining with \six\ yielded almost indistinguishable results.


\label{lastpage}
\end{document}